\documentclass[twoside,twocolumn,9pt]{article}
\usepackage{extsizes}
\usepackage[super,sort&compress,comma]{natbib} 
\usepackage[version=3]{mhchem}
\usepackage[left=1.5cm, right=1.5cm, top=1.785cm, bottom=2.0cm]{geometry}
\usepackage{balance}
\usepackage{mathptmx}
\usepackage{sectsty}
\usepackage{graphicx} 
\usepackage{lastpage}
\usepackage[format=plain,justification=justified,singlelinecheck=false,font={stretch=1.125,small,sf},labelfont=bf,labelsep=space]{caption}
\usepackage{float}
\usepackage{fancyhdr}
\usepackage{fnpos}
\usepackage[english]{babel}
\addto{\captionsenglish}{%
  
}
\usepackage{array}
\usepackage{droidsans}
\usepackage{charter}
\usepackage[T1]{fontenc}
\usepackage[usenames,dvipsnames]{xcolor}
\usepackage{setspace}
\usepackage[compact]{titlesec}
\usepackage{hyperref}

\usepackage{epstopdf}

\definecolor{cream}{RGB}{222,217,201}

\begin{document}

\pagestyle{fancy}
\thispagestyle{plain}
\fancypagestyle{plain}{
\renewcommand{\headrulewidth}{0pt}
}

\newcommand{\bra}[1]{\langle{#1}|}
\newcommand{\ket}[1]{|{#1}\rangle}
\newcommand{\tbra}{\texttt{bra} }
\newcommand{\tket}{\texttt{ket} }
\newcommand{\beq}{\begin{equation}}
\newcommand{\eeq}{\end{equation}}
\newcommand{\half}{\frac{1}{2}}
\newcommand{\quart}{\frac{1}{4}}
\newcommand{\op}{\hat}
\newcommand{\mat}[1]{\mathbf{#1}}
\renewcommand{\vec}[1]{{\mathbf{#1}}}
\newcommand{\eq}[1]{Eq.~(\ref{#1})}
\newcommand{\eqs}[2]{Eqs.~(\ref{#1},~\ref{#2})}
\newcommand{\sect}[1]{Section~\ref{#1}}
\newcommand{\fig}[1]{Figure~\ref{#1}}
\newcommand{\tab}[1]{Table~\ref{#1}}
\newcommand{\dg}{\dagger}
\newcommand{\lk}{\left}
\newcommand{\rk}{\right}
\newcommand{\onlinecite}{\citenum}
\newcommand{\chg}[1]{{\color{blue} #1}}

\makeFNbottom
\makeatletter
\renewcommand\LARGE{\@setfontsize\LARGE{15pt}{17}}
\renewcommand\Large{\@setfontsize\Large{12pt}{14}}
\renewcommand\large{\@setfontsize\large{10pt}{12}}
\renewcommand\footnotesize{\@setfontsize\footnotesize{7pt}{10}}
\makeatother

\renewcommand{\thefootnote}{\fnsymbol{footnote}}
\renewcommand\footnoterule{\vspace*{1pt}%
\color{cream}\hrule width 3.5in height 0.4pt \color{black}\vspace*{5pt}} 
\setcounter{secnumdepth}{5}

\makeatletter 
\renewcommand\@biblabel[1]{#1}            
\renewcommand\@makefntext[1]%
{\noindent\makebox[0pt][r]{\@thefnmark\,}#1}
\makeatother 
\renewcommand{\figurename}{\small{Fig.}~}
\sectionfont{\sffamily\Large}
\subsectionfont{\normalsize}
\subsubsectionfont{\bf}
\setstretch{1.125} 
\setlength{\skip\footins}{0.8cm}
\setlength{\footnotesep}{0.25cm}
\setlength{\jot}{10pt}
\titlespacing*{\section}{0pt}{4pt}{4pt}
\titlespacing*{\subsection}{0pt}{15pt}{1pt}

\fancyfoot{}
\fancyfoot[LO,RE]{\vspace{-7.1pt}\includegraphics[height=9pt]{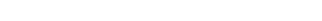}}
\fancyfoot[CO]{\vspace{-7.1pt}\hspace{13.2cm}\includegraphics{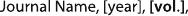}}
\fancyfoot[CE]{\vspace{-7.2pt}\hspace{-14.2cm}\includegraphics{head_foot/RF}}
\fancyfoot[RO]{\footnotesize{\sffamily{1--\pageref{LastPage} ~\textbar  \hspace{2pt}\thepage}}}
\fancyfoot[LE]{\footnotesize{\sffamily{\thepage~\textbar\hspace{3.45cm} 1--\pageref{LastPage}}}}
\fancyhead{}
\renewcommand{\headrulewidth}{0pt} 
\renewcommand{\footrulewidth}{0pt}
\setlength{\arrayrulewidth}{1pt}
\setlength{\columnsep}{6.5mm}
\setlength\bibsep{1pt}

\makeatletter 
\newlength{\figrulesep} 
\setlength{\figrulesep}{0.5\textfloatsep} 

\newcommand{\topfigrule}{\vspace*{-1pt}%
\noindent{\color{cream}\rule[-\figrulesep]{\columnwidth}{1.5pt}} }

\newcommand{\botfigrule}{\vspace*{-2pt}%
\noindent{\color{cream}\rule[\figrulesep]{\columnwidth}{1.5pt}} }

\newcommand{\dblfigrule}{\vspace*{-1pt}%
\noindent{\color{cream}\rule[-\figrulesep]{\textwidth}{1.5pt}} }

\makeatother

\twocolumn[
  \begin{@twocolumnfalse}
{\includegraphics[height=30pt]{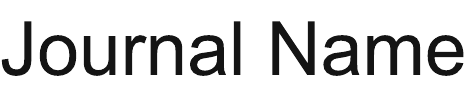}\hfill\raisebox{0pt}[0pt][0pt]{\includegraphics[height=55pt]{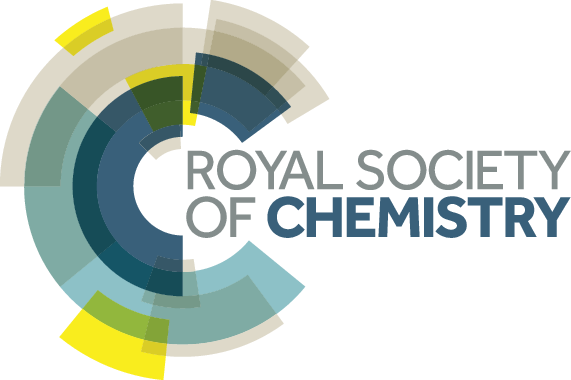}}\\[1ex]
\includegraphics[width=18.5cm]{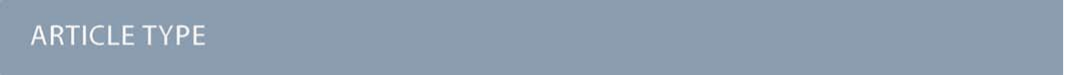}}\par
\vspace{1em}
\sffamily
\begin{tabular}{m{4.5cm} p{13.5cm} }

\includegraphics{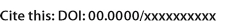} & \noindent\LARGE{\textbf{Orbital optimisation in xTC transcorrelated methods}} \\
\vspace{0.3cm} & \vspace{0.3cm} \\

 & \noindent\large{Daniel Kats,$^{\ast}$ Evelin M. C. Christlmaier,
Thomas Schraivogel, and Ali Alavi} \\

\includegraphics{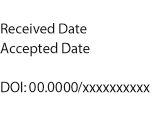} & \noindent\normalsize{
We present a combination of the bi-orthogonal orbital optimisation framework with 
the recently introduced xTC version of transcorrelation. 
This allows us to implement non-iterative perturbation based methods on top of the 
transcorrelated Hamiltonian. 
Besides, the orbital optimisation influences results of other truncated methods,
such as the distinguishable cluster with singles and doubles. 
The accuracy of these methods in comparison to standard xTC methods is demonstrated, 
and the advantages and disadvantages of the orbital optimisation are discussed.
} \\

\end{tabular}

 \end{@twocolumnfalse} \vspace{0.6cm}

  ]

\renewcommand*\rmdefault{bch}\normalfont\upshape
\rmfamily
\section*{}
\vspace{-1cm}


\footnotetext{\textit{Max Planck Institute for Solid State Research, Heisenbergstr. 1, 70569 Stuttgart, Germany}}
\footnotetext{\textit{$^{\ast}$~E-Mail: d.kats@fkf.mpg.de}}


\section{Introduction}
An accurate description of the electron correlation is crucial for the understanding of many 
chemical and physical phenomena.
Coupled cluster (CC) methods \cite{cizek:66} are among the most accurate and widely used 
wavefunction-based methods to describe the electron correlation, 
and are often considered as the gold standard for the description of the dynamical electron 
correlation in molecular systems.
However, the computational cost of the CC methods scales steeply with the system size
with increasing the excitation level of the cluster operator, and therefore in practice
the CC methods are often limited to the singles and doubles excitations (CCSD), 
and the triples corrections have to be added perturbatively (CCSD(T)).
Linear scaling implementations of the CC methods have been developed, 
\cite{Hampel:96,Schuetz:00b,Werner:2011a,Riplinger:2016,Schmitz:2016,Schwilk:2015,Schwilk:2017,Ma:2017a}
but the complexity and larger computational-cost prefactor of the linear scaling algorithms 
still limits the underlying CC methods to CCSD(T).

An alternative approach to improve the accuracy of CC methods without going to high excitations
is to modify the amplitude equations. 
\cite{Meyer:71,paldus_approximate_1984,piecuch_solution_1991,piecuch_approximate_1996,kowalski_method_2000,
bartlett_addition_2006,nooijen_orbital_2006, Neese:09, huntington_pccsd:_2010,robinson_approximate_2011,
huntington_accurate_2012,paldusExternally2017,black_statistical_2018,behnleREMP2019,behnleOOREMP2021,
 kats_dc_2013,kats_dcsd_2014,kats_accurate_2015,katsDistinguishable2019,rishiCan2019}
The distinguishable cluster singles and doubles (DCSD) approach \cite{kats_dc_2013,kats_dcsd_2014}
is one of such approaches, and has been shown in numerous benchmark calculations
to be more accurate than the standard CCSD method. 
\cite{kats_accurate_2015,kats_improving_2018,rishi_excited_2017,tsatsoulis_comparison_2017,
black_statistical_2018,limanniRole2019,vitaleFCIQMCTailored2020,linFragmentBased2020,schraivogelAccuracy2021}

Another well-known issue of the wavefunction-based electron-correlation methods is the 
requirement of large basis sets to achieve high accuracy. Quadruple- or even pentuple-zeta
basis sets are often required to achieve the chemical accuracy in the calculations of
relative energies using CCSD(T) or higher order methods.
Introducing explicit correlation into the wavefunction, \textit{i.e.}, functions which
explicitly depend on the electron-electron distances, is a way to reduce the basis set 
incompleteness error and to improve the accuracy of the results.
An established approach for coupled-cluster type methods to introduce the explicit correlation 
is the F12 method, \cite{k1985,kk1991i,ks2002,m2003,tenno2004,tenno2004a,v2004,tk2005,kedzuch:05,fkh05,fhk06b,f12g,noga:07,ccf12,
tknh07,rmp2f12,skhv08a,skhv08b,nkst08,tkh2008,valeev08,valeev08b,Torheyden:2008,bokhan2008,kaw2009,Werner:2011,
Ten-no:TCA131-1,Hattig:CR112-4,Kong:CR112-75,tewRelaxing2018,katsOrbitalOptimized2019} 
which has been shown to be very accurate and efficient 
in many calculations, and has also been extended to other wavefunction-based methods, 
\cite{shiozaki_multireference_2013,vitaleFCIQMCTailored2020}
\textit{e.g.}, the full configuration interaction quantum Monte Carlo (FCIQMC) method, \cite{boothExplicitly2012}
linear scaling methods, \cite{Manby:06a,Tew:2011,Ma:2017,Pavosevic:2017} and periodic systems. 
\cite{usvyat_linear-scaling_2013,gruneisEfficient2015}
Despite tremendous success of the F12 method, it is not without its limitations.
It requires new auxiliary basis sets, involves various additional approximations, 
and it is very hard and computationally expensive to extend beyond the single and double 
excitations level. \cite{Koehn:2009} 

An alternative approach to introduce the explicit correlation is transcorrelation,
\cite{boysCalculation1969,ten-noFeasible2000,htt2001,hinoApplication2002,
umezawa_transcorrelated_2003,
yanai_canonical_2006,yanai_canonical_2007,yanai_canonical_2012,
tsuneyukiTranscorrelated2008,
ochi_efficient_2012,ochi_optical_2014,ochi_second-order_2015,ochi_iterative_2016,wahlen-strothman_lie_2015,luoCombining2018,
dobrautzCompact2019,cohenSimilarity2019,baiardi_transcorrelated_2020,khamoshi_exploring_2021,
giner_new_2021,gutherberyliumdimer2021,liaoEfficient2021,liao22,
hauptOptimizing2023,christlmaierXTC2023,
ammarBiorthonormal2023,leeStudies2023,ammarTranscorrelated2023}
which is based on a similarity transformation of the Hamiltonian using a pre-optimised Jastrow 
factor. Transcorrelation has been shown to not only reduce the basis set incompleteness
error, but also to improve the accuracy of the wavefunction-based methods employed to solve the
transcorrelated Schr\"odinger equation.  
\cite{liaoEfficient2021,schraivogelTranscorrelated2023,hauptOptimizing2023,christlmaierXTC2023}
Transcorrelation has been combined with the CCSD and DCSD methods (TC-CCSD and TC-DCSD), 
and has been shown to be more accurate than the CCSD-F12 and DCSD-F12 methods. 
\cite{TCCC2021,schraivogelTranscorrelated2023,christlmaierXTC2023}
Especially the TC-DCSD method yields very accurate results for the relative energies of
atoms and molecular systems, with accuracy approaching CCSD(T)-F12. 
\cite{TCCC2021,schraivogelTranscorrelated2023}
This requires well-optimised Jastrow factors, and the optimisation of the Jastrow factor
in these studies has been done by minimising the variance of the reference energy, 
\cite{hauptOptimizing2023,christlmaierXTC2023} using variational Monte-Carlo.
One of the main advantages of the transcorrelation is that it allows to apply almost any
standard wavefunction-based method to the transcorrelated Hamiltonian. 
However, the similarity transformation of the Hamiltonian using the Jastrow factor
results in a non-Hermitian Hamiltonian with a non-diagonal Fock matrix, and the standard 
non-iterative perturbative methods based on the M{\o}ller-Plesset partitioning of the 
Hamiltonian,
such as MP2 or CCSD(T), are not directly applicable to the transcorrelated Hamiltonian.

Another issue of the transcorrelated Hamiltonian are the three-electron integrals,
which are computationally expensive and inconvenient for the implementation of  
wavefunction-based methods.
Recently, we have introduced an approximation to the transcorrelation -- the xTC approach -- 
that allows to
neglect the explicit three-electron integrals in the transcorrelated Hamiltonian by 
incorporating the three-electron terms into the zero-, one-, and two-electron integrals,
\cite{christlmaierXTC2023}
which barely affects the accuracy of the transcorrelated calculations, and substantially
reduces the computational cost and scaling of the method.

The orbital optimisation is a crucial part of the truncated wavefunction-based methods, 
and can improve the accuracy of the methods. 
Additionally, the Hartree-Fock type orbital optimisation leads to a diagonal Fock matrix,
which is a key ingredient for the non-iterative perturbative M{\o}ller-Plesset methods. 
The non-Hermitian nature of the transcorrelated Hamiltonian prevents the standard
methods to optimise the orbitals, and the biorthogonal orbital optimisation has to be employed.
\cite{htt2001,ammarBiorthonormal2023,leeStudies2023,ammarTranscorrelated2023}
In this work, we present a combination of the bi-orthogonal orbital optimisation framework
with the xTC version of the transcorrelation, and demonstrate the accuracy of the
non-iterative perturbation based methods on top of the transcorrelated Hamiltonian,
and the effect of the orbital optimisation on the results of other truncated methods.

\section{Theory}
\subsection{xTC Transcorrelation}
\label{sec:xtc}
In this section we briefly review the transcorrelation, especially the optimisation 
of the Jastrow factor, and the xTC methods.
The full details on these methods can be found in 
Refs.~\onlinecite{hauptOptimizing2023,christlmaierXTC2023}.

The transcorrelation is based on a similarity transformation of the Hamiltonian
using a pre-optimised Jastrow factor, which accounts for a portion of the electron correlation,
\begin{equation}
  \tilde H = e^{-\tau} \hat H e^{\tau}.
\end{equation}
The resulting transcorrelated Hamiltonian $\tilde H$ is non-Hermitian, 
and can be inserted into the Schr\"odinger equation instead of the standard 
Hamiltonian, which effectively factorises the total wavefunction into the
Jastrow factor contribution and the rest, 
\begin{equation}
  \tilde H \Psi = e^{-\tau} \hat H e^{\tau} \Psi = E \Psi.
\end{equation}
If $\tau$ is defined as a sum of pair-wise correlation operators,
\begin{equation}
  \tau = \sum_{i<j} u(\vec r_i, \vec r_j),
\end{equation}
with $u(\vec r_i, \vec r_j)$ being a function of the coordinates of two electrons,
the transcorrelated Hamiltonian can be written as a sum of one-, two- and three-electron 
operators,
\begin{equation}
  \label{eq:tch}
  \begin{aligned}
  \tilde H &= \hat H + [\hat H, \tau] + \half[[\hat H, \tau], \tau] + 
  \frac{1}{6} [[[ \hat H, \tau], \tau], \tau]\\
  &= E_{\rm nuc} + h_p^q a^\dg_{p\sigma}a_{q \sigma} +
\half \left(V_{pr}^{qs} - K_{pr}^{qs}\right)a^\dg_{p\sigma} a^\dg_{r\rho} a_{s \rho} a_{q \sigma}\\
&-\frac{1}{6}L_{prt}^{qsu}a^\dg_{p\sigma} a^\dg_{r\rho}a^\dg_{t\tau} a_{u \tau} a_{s \rho} a_{q \sigma}.
  \end{aligned}
\end{equation}
Here and in the following, we use the Einstein summation convention, 
and the indices $p, q, r, s, t, u$ denote the general spatial orbitals and
$\sigma, \rho, \tau$ the spin.
$h_p^q$ is the one-electron part of the Hamiltonian, 
$V_{pr}^{qs}$ and $K_{pr}^{qs}$ are the two-electron integrals (with $K$ being the additional 
term due to the transcorrelation), and $L_{prt}^{qsu}$ is the three-electron integral.
In principle, $\tau$ also contains one-electron terms, but in our current implementation
these terms are added to the two-electron functions $u(\vec r_i, \vec r_j)$. 

The Jastrow factor optimisation is a crucial part of the transcorrelation method.
In our recent works, \cite{schraivogelTranscorrelated2023,hauptOptimizing2023,christlmaierXTC2023} we have demonstrated
that the optimisation based on the 
minimisation of the variance of the reference energy,
\begin{equation}
  \label{eq:variance}
\sigma^2_{\rm ref} = \bra{\Phi_0} e^{-\tau} \left(\hat H - E_{\rm ref}\right)^2 e^{\tau} 
\ket{\Phi_0},
\end{equation}
yields Jastrow factors which
not only reduce the basis set incompleteness error, but also improve the accuracy
of the wavefunction-based methods employed to solve the transcorrelated Schr\"odinger equation. 
This can be easily understood by inserting $e^{\tau}e^{-\tau}$ and the resolution of the 
identity into expression for the variance, 
\eq{eq:variance}, which yields
\begin{equation}
  \label{eq:variance2}
  \sigma^2_{\rm ref} = \sum_{i\ne 0}\bra{\Phi_0} \tilde H \ket{\Phi_i}\bra{\Phi_i} \tilde H 
  \ket{\Phi_0},
\end{equation}
where we have utilised the definition of the transcorrelated reference energy $E_{\rm ref}$ 
as the expectation value of the transcorrelated Hamiltonian
with respect to the reference determinant $\Phi_0$,
\begin{equation}
  \label{eq:eref}
  E_{\rm ref} = \bra{\Phi_0} \tilde H \ket{\Phi_0}.
\end{equation}
Thus, the variance of the reference energy is a measure of the electron correlation 
not accounted for by the Jastrow ansatz, $e^{\tau}\Phi_0$, 
and the smaller the variance, the less electron correlation has to be accounted for by the 
wavefunction-based methods.
In practice, the variance of the reference energy is minimised using the variational 
Monte-Carlo (VMC) method.

The three-electron integrals in the transcorrelated Hamiltonian, \eq{eq:tch}, 
are inconvenient for the implementation of the wavefunction-based methods.
They are not only computationally expensive to evaluate, 
but also introduce a large number of new terms in the wavefunction-based 
methods, e.g., coupled cluster methods,
and increase the computational scaling of the method. 
\cite{TCCC2021,schraivogelTranscorrelated2023}
We have demonstrated that the explicit three-electron integrals can be neglected
by incorporating the three-electron terms into the zero-, one-, and two-electron 
integrals through the normal ordering of the transcorrelated Hamiltonian with respect
to the reference determinant, \cite{TCCC2021,schraivogelTranscorrelated2023,christlmaierXTC2023}
which barely affects the accuracy of the transcorrelated calculations.
Recently, we have developed and implemented a strategy to efficiently evaluate the
three-electron-contributions which takes advantage of the grid-based computation 
of the transcorrelated integrals, and allows to calculate the modified two-electron
(and lower) integrals on the fly. \cite{christlmaierXTC2023}
As a result, the nominal computational scaling of the evaluation of the transcorrelated 
integrals is reduced from $\mathcal{O}(N^7)$ to $\mathcal{O}(N^5)$, 
where $N$ corresponds to the size of the molecular system.
Besides, the new Hamiltonian contains only zero-, one-, and two-electron terms, 
\begin{equation}
\label{eq:xtch}
\tilde H = \tilde E_{\rm nuc} + \tilde h_p^q a^\dg_{p\sigma}a_{q \sigma} +
\half \tilde V_{pr}^{qs} a^\dg_{p\sigma} a^\dg_{r\rho} a_{s \rho} a_{q \sigma},
\end{equation}
and therefore almost any standard wavefunction-based methods can be applied -- 
as long as the non-Hermitian nature of the transcorrelated Hamiltonian is taken into account --
and the computational scaling of the method remains the same as for the standard Hamiltonian.
We have termed this approach the xTC method, and have demonstrated its accuracy
in a combination with CCSD, DCSD and CCSDT methods
for various chemical systems. \cite{christlmaierXTC2023}

We note in passing that   
since the normal ordering for open-shell systems is spin-dependent, 
the xTC integrals are also spin-dependent, even if the reference determinant is spin-restricted.
However, our xTC implementation is flexible with respect to the choice of the reference 
determinant, since it does not rely on the diagonality of the 1-body reduced density matrix 
(1-RDM). This allows for example to use the xTC approach with the correlated 1-RDMs, which
can be obtained in a preceding coupled cluster calculation; or one can also use spin-averaged 
1-RDMs, which leads to spin-independent xTC integrals (for a restricted reference determinant),
and our benchmark calculations have shown that the spin-independent xTC integrals are
as accurate as the spin-dependent ones. \cite{christlmaierXTC2023}

\subsection{Biorthogonal Orbital Optimisation}
The integrals in the xTC approach are computed in the basis of the molecular orbitals
from the reference determinant, which is obtained from a mean-field calculation \textit{before}
the transcorrelation, and the orbitals are not changed in the subsequent wavefunction-based
calculation.
Thus, the orbitals are not optimised for the transcorrelated Hamiltonian, and the 
final accuracy might by improved by reoptimising the orbitals.
However, the transcorrelated Hamiltonian is non-Hermitian, and the standard
methods to optimise the orbitals, such as the Hartree-Fock method, are not directly applicable.
Instead, one has to employ the biorthogonal orbital optimisation, in which the \tbra and \tket
orbitals are different and represent two mutually orthonormal sets,
\begin{equation}
  \label{eq:biorth}
  \bra{\bar\phi_p} \phi_q\rangle = \delta_{pq}, 
\end{equation}
where $\bra{\bar\phi_p}$ and $\ket{\phi_q}$ are the \tbra and \tket orbitals, respectively.
The orbital coefficients are obtained by minimising the reference energy, \eq{eq:eref},
with respect to the \tbra and \tket orbitals with the biorthogonality constraint,
\eq{eq:biorth}.
This is achieved by solving the coupled self-consistent field (SCF) equations
(for simplicity, we show only the closed-shell case and assume the orthogonality of the 
original orbitals),
\begin{equation}
  \label{eq:biorthscf}
    \mat F \mat C = \mat C \epsilon, \quad
    \mat {\bar C}^\dg \mat F = \epsilon \mat {\bar C}^\dg,
\end{equation}
where $\mat F$ is the Fock matrix, 
\begin{equation}
  \label{eq:fock}
  \begin{aligned}
  F_{p}^{q} &= \tilde h_{p}^{q} + \gamma^r_s \left(\tilde V_{pr}^{qs} - \half \tilde V_{pr}^{sq}\right),\\
  \gamma^r_s &= 2\sum_{i \in \rm occ} \bar C^{\dg r}_{i} C_s^i ,
  \end{aligned}
\end{equation}
$\mat C$ and $\mat {\bar C}$ are \tbra and \tket coefficient matrices which transform
from the previous molecular orbitals to the new ones, and $\epsilon$ is the diagonal
matrix of orbital energies. 
Here and in the following, $i, j, k, l, \ldots$ denote the occupied orbitals, 
and $a, b, c, d, \ldots$ the virtual orbitals.
The \tbra and \tket coefficient matrices are interconnected through the biorthogonality condition,
\eq{eq:biorth},
and therefore the conjugate transpose of the \tbra matrix is the inverse of the \tket matrix, 
$\mat {\bar C}^\dg = \mat C^{-1}$.
The equations for a biorthogonal unrestricted Hartree-Fock 
method can be obtained in a similar fashion.

The \eq{eq:biorthscf} is solved iteratively until the change in the orbitals is small enough.
In principle, the calculation of the Fock matrix, \eq{eq:fock}, requires recalculation
of the xTC integrals in every iteration, since the change in the 1-RDM affects the
xTC approximation, but in practice we assume that the change in the 1-RDM is small, thus, 
the effect of this onto the xTC integrals can be neglected, which
immensely reduces the computational cost of the biorthogonal orbital optimisation.
Hence, the integrals are calculated only once, in the original molecular orbital basis, 
and the Fock matrix is updated in every iteration using new coefficient matrices 
$\mat C$ and $\mat {\bar C}$ according to \eq{eq:fock}.

Standard techniques to optimise the orbitals, such as the direct inversion of the iterative
subspace (DIIS) method, can be employed to accelerate the convergence of the biorthogonal SCF.

Since the Fock matrices are non-Hermitian, the orbital optimisation is not guaranteed to 
yield real orbitals and orbital energies, and in practice small imaginary 
parts of the orbital energies are observed.
However, in our experience, the imaginary parts of the orbital energies are very small,
and occur only rarely and only for the virtual orbitals, and therefore the density matrices
and the Fock matrices remain real, and the SCF equations can be solved using real algebra.
For the correlated calculations, the complex-valued orbital coefficients are
transformed into the real-valued ones by identifying complex-conjugated pairs of the 
orbital energies and using the (normalized) real and imaginary parts of the corresponding 
orbital coefficients as the new orbital coefficients.
As a result, the final orbitals are real, and the Fock matrix is diagonal 
(apart from the $2\times 2$ blocks which correspond to rotated orbitals),
and the wavefunction-based methods can be applied to the real-valued xTC Hamiltonian.

The optimisation of the orbitals in the xTC method changes the reference determinant
for the subsequent wavefunction-based methods, and therefore the Jastrow factor is no longer
optimal for the new reference determinant, cf. \eq{eq:variance2}. 
This can be remedied by reoptimising the Jastrow factor, but this would require a VMC 
calculation with different \tbra and \tket orbitals, which is not straightforward to implement.
Therefore, we have not reoptimised the Jastrow factor in the present work, and thus
the transcorrelated results can actually deteriorate after the orbital optimisation.

As an alternative to the biorthogonal orbital optimisation, one can employ a biorthogonal 
\textit{pseudo-canonicalisation} of the orbitals, which is a non-iterative method to obtain
diagonal \textit{blocks} of the Fock matrix in the occupied and virtual orbital subspaces,
and which does not change the reference determinant.
For this purpose, the transcorrelated Fock matrix is constructed in the original molecular orbital basis
according to \eq{eq:fock}, and then diagonalised in the occupied and virtual subspaces,
which yields the new orbital coefficients. If complex-valued orbital coefficients occur,
the real-valued orbitals are obtained as described above.
This procedure does not change the final energy of the non-perturbative methods, e.g., CCSD
or DCSD, but allows to apply the perturbative (M{\o}ller-Plesset) methods , e.g., MP2 or CCSD(T),
on top of the xTC Hamiltonian, and to obtain the perturbative corrections to the energy.
The results of the perturbative methods calculated with the biorthogonal pseudo-canonical
orbitals are exactly the same as the results one would obtain with iterative calculations
of the perturbative corrections, but the computational cost is substantially reduced. 
The only source of deviation are the complex eigenvalues of the Fock matrix, which are
however very rare and have a very small imaginary part, and therefore the effect of these
deviations on the final results is negligible.
Note that the occupied-virtual and virtual-occupied blocks of the final Fock matrix
are not zero in the pseudo-canonical case, and therefore the perturbative methods should include 
corrections involving these blocks. 

\subsection{xTC Coupled Cluster/Perturbative Methods}
\label{sec:cc}
Coupled cluster methods are based on the exponential ansatz for the wavefunction,
\begin{equation}
  \label{eq:ccwf}
  \ket{\Psi} = e^{\hat T} \ket{\Phi_0},
\end{equation}
where $\ket{\Phi_0}$ is the reference determinant, and $\hat T$ is the cluster operator,
which is a sum of excitation operators,
\begin{equation}
  \label{eq:ccop}
  \hat T = \hat T_1 + \hat T_2 + \ldots,
\end{equation}
where $\hat T_n$ is the $n$-electron excitation operator.
If the cluster operator is truncated at the two-electron level, the method is termed
CCSD. The cluster operator is determined by solving the amplitude equations,
which are obtained by inserting the exponential ansatz, \eq{eq:ccwf}, into the Schr\"odinger
equation, and projecting onto the excited determinants. 
In the distinguishable cluster approach the amplitude equations are slightly different,
but the computational scaling and the efficiency of the method are the same 
as for the standard coupled cluster methods (or slightly better).
As mentioned above,
if the biorthogonal orbital optimisation is employed, the reference determinant $\Phi_0$
in \eq{eq:ccwf} is not the same as the original reference determinant in the Jastrow
optimisation, \eqs{eq:variance}{eq:variance2}.

The xTC Hamiltonian, \eq{eq:xtch}, contains only upto two-electron terms, and therefore
standard coupled cluster amplitude equations can be used to solve the transcorrelated
Schr\"odinger equation. The only difference to the standard coupled cluster implementations 
is the non-Hermitian nature of the xTC Hamiltonian, \textit{i.e.}, 
$\tilde V_{pr}^{qs} \ne \tilde V_{qs}^{pr}$, and (in general) a non-diagonal Fock matrix, but
this does not affect the computational scaling or efficiency of the method.
The explicit amplitude equations for closed-shell CCSD and DCSD, and the unrestricted 
versions (UCCSD and UDCSD) as implemented in the 
\texttt{ElemCo.jl} package \cite{elemcojl} can be found in the documentation of the 
package. \cite{elemcojl-docs}

The perturbative methods based on the M{\o}ller-Plesset partitioning of the Hamiltonian
can also be applied to the xTC Hamiltonian, however, if the Fock matrix is non-diagonal,
the perturbative corrections have to be calculated iteratively, which substantially
increases the computational cost of the method, \textit{e.g.}, in the case of CCSD(T)
one would have to store and iterate the triples amplitudes.
The biorthogonal optimisation ensures that the Fock matrix is diagonal 
(up to the occasional $2\times 2$ blocks in the virtual space, \textit{vide supra}), 
and therefore the perturbative corrections can be calculated non-iteratively.
The MP2 correlation energy can be obtained by the standard formula 
(taking into account the non-Hermitian nature of the xTC Hamiltonian), 
\textit{e.g.}, in the closed-shell case,
\begin{equation}
\label{eq:xtc-mp2}
  E_{\textrm{xTC-MP2}} = \frac{\left(2 \tilde V_{ij}^{ab}-\tilde V_{ij}^{ba}\right) \tilde V_{ab}^{ij}}
  {\epsilon_i + \epsilon_j - \epsilon_a - \epsilon_b} 
  + 2\frac{F_{i}^{a} F_{a}^{i}}{\epsilon_i - \epsilon_a}.
\end{equation}
The second term is important for the pseudo-canonical orbitals, and is zero for the 
fully optimised orbitals.

The combinations of the perturbative triples correction in CCSD(T) with the xTC method
is more complicated, since it formally involves 
singles and doubles amplitudes corresponding to \tbra and \tket wavefunctions,
\textit{e.g.}, in the closed-shell formalism,
\begin{equation}
\label{eq:Eccsd(t)}
\begin{aligned}
&E_{(T)} =E_{[T]} + \sum_{i\le j\le k}p(i,j,k) \left[
  \tilde V_{jk}^{bc} X_{abc}^{ijk} T_{i}^{\dagger a} 
  + \tilde V_{ik}^{ac} X_{abc}^{ijk} T_{j}^{\dagger b}\right.\\
&+ \tilde V_{ij}^{ab} X_{abc}^{ijk} T_{k}^{\dagger c} 
  + T_{jk}^{\dagger bc} X_{abc}^{ijk} F_i^a  
\left. + T_{ik}^{\dagger ac} X_{abc}^{ijk} F_j^b
+ T_{ij}^{\dagger ab} X_{abc}^{ijk} F_k^c \right],
\end{aligned}
\end{equation}
where $E_{[T]}$ is the [T]-triples correction to the energy, 
\begin{equation}
E_{[T]} = \sum_{i\le j\le k}p(i,j,k) K^{abc}_{ijk} X_{abc}^{ijk},
\end{equation}
$p(i,j,k)$ are prefactors which account for the triangular summation,
\begin{equation}
p(i,j,k) = \left[\begin{matrix}
  2 & i\ne j\ne k\\
  1 & i=j \oplus j=k\\
  0 & i=j=k
\end{matrix}\right.\\
\end{equation}
and $X_{abc}^{ijk}$, $K_{abc}^{ijk}$ and $K^{abc}_{ijk}$ correspond to the contravariant triples
amplitudes, 
\begin{equation}
X_{abc}^{ijk} = \frac{4K_{abc}^{ijk} - 2K_{acb}^{ijk} - 2K_{cba}^{ijk} - 2K_{bac}^{ijk}
+ K_{cab}^{ijk} + K_{bca}^{ijk}}
{\epsilon_i +\epsilon_j +\epsilon_k -\epsilon_a -\epsilon_b -\epsilon_c},
\end{equation}
the right-hand side of the triples equations, 
\begin{equation}
\begin{aligned}
K_{abc}^{ijk} &= \tilde V_{bc}^{dk} T^{ij}_{ad} + \tilde V_{ac}^{dk} T^{ij}_{db} 
+ \tilde V_{cb}^{dj} T^{ik}_{ad} + \tilde V_{ab}^{dj} T^{ik}_{dc} \\
&+ \tilde V_{ca}^{di} T^{jk}_{bd} + \tilde V_{ba}^{di} T^{jk}_{dc} 
- \tilde V_{lc}^{jk} T^{li}_{ba} - \tilde V_{lc}^{ik} T^{lj}_{ab} \\
&- \tilde V_{lb}^{kj} T^{li}_{ca} - \tilde V_{lb}^{ij} T^{lk}_{ac} 
- \tilde V_{la}^{ki} T^{lj}_{cb} - \tilde V_{la}^{ji} T^{lk}_{bc}, 
\end{aligned}
\end{equation}
and its \tbra counterpart,
\begin{equation}
\label{eq:Kabc}
\begin{aligned}
K^{abc}_{ijk} &= \tilde V_{dk}^{bc} T_{ij}^{\dg ad} + \tilde V_{dk}^{ac} T_{ij}^{\dg db}
+ \tilde V_{dj}^{cb} T_{ik}^{\dg ad} + \tilde V_{dj}^{ab} T_{ik}^{\dg dc} \\
&+ \tilde V_{di}^{ca} T_{jk}^{\dg bd} + \tilde V_{di}^{ba} T_{jk}^{\dg dc} 
- \tilde V_{jk}^{lc} T_{li}^{\dg ba} - \tilde V_{ik}^{lc} T_{lj}^{\dg ab} \\
&- \tilde V_{kj}^{lb} T_{li}^{\dg ca} - \tilde V_{ij}^{lb} T_{lk}^{\dg ac} 
- \tilde V_{ki}^{la} T_{lj}^{\dg cb} - \tilde V_{ji}^{la} T_{lk}^{\dg bc}.
\end{aligned}
\end{equation}
The conventional replacement of the \tbra amplitudes by the \tket amplitudes 
is theoretically less justified in the case of the xTC Hamiltonian,
because of the non-Hermitian nature of the Hamiltonian.
Besides, the integrals involved in the calculation of $X_{abc}^{ijk}$ and $K^{abc}_{ijk}$ 
are different, and therefore one cannot simply replace $K^{abc}_{ijk}$ by $K_{abc}^{ijk}$
as in the standard CCSD(T) method.
Thus, instead of the standard CCSD(T) method, we have employed the 
$\Lambda$CCSD(T) method,\cite{taubeImproving2008} which is very similar to the standard CCSD(T) method,
but the \tbra amplitudes in \eqs{eq:Eccsd(t)}{eq:Kabc} are replaced by
Lagrange multipliers; in the closed-shell formalism -- covariant Lagrange multipliers, 
\begin{equation}
\begin{aligned}
\bar \Lambda_{ij}^{ab} &= \frac{2}{3}\Lambda_{ij}^{ab} + \frac{1}{3} \Lambda_{ij}^{ba},\\
\bar \Lambda_i^a &= \frac{1}{2} \Lambda_i^a.
\end{aligned}
\end{equation}
The Lagrange multiplier equations for closed-shell and for unrestricted formalism
can be found in the documentation of the \texttt{ElemCo.jl} package. \cite{elemcojl-docs}

In the following xTC-BO-MP2 and xTC-BO-$\Lambda$CCSD(T) denote the xTC methods 
based on the optimized biorthogonal orbitals, and xTC-pcBO-MP2 and xTC-pcBO-$\Lambda$CCSD(T)
the ones based on the pseudo-canonical biorthogonal orbitals. 
For the sake of brevity, we will refer to these perturbative methods as xTC-MP2 and xTC-CCSD(T).

\section{Computational Details}

The closed-shell and unrestricted versions of the biorthogonal orbital Hartree-Fock, 
the pseudo-canonicalisation, and the coupled cluster methods from \sect{sec:cc} 
were implemented in the \texttt{ElemCo.jl} package. \cite{elemcojl}

We utilise the Drummond-Towler-Needs form \cite{drummond_jastrow_2004} of $u(\vec r_i, \vec r_j)$ in the Jastrow factors, 
\begin{equation}
  u(\vec r_i, \vec r_j) = v(r_{ij}) +  
  \sum_I \frac{1}{N_{\textrm{el}} - 1} \left[\chi(r_{iI}) + \chi(r_{jI})\right] 
  + f(r_{ij}, r_{iI}, r_{jI}),
\end{equation}
which includes terms for electron-electron ($v$), electron-nucleus ($\chi$), 
and electron-electron-nucleus ($f$) interactions, expanded in natural powers. 
The Jastrow factors have been optimised using VMC in the \texttt{CASINO} package \cite{needs20} 
by minimising the reference energy variance as described in \sect{sec:xtc} and in more details in
Ref.~\onlinecite{hauptOptimizing2023}. 

The xTC contributions to the integrals were calculated numerically in the 
\texttt{TCHINT} program, \cite{tchint}
and added to the standard integrals obtained from the \texttt{MOLPRO} package. \cite{MOLPRO}
For the numerical integration, we used atom-centered grids formed from Treutler-Ahlrichs 
radial grids and Lebedev angular grids obtained from \texttt{PySCF} \cite{PySCF} (grid level 2).
The transcorrelated integrals are then used in the coupled-cluster calculations in the
\texttt{ElemCo.jl} package through a FCIDUMP interface. 

The benchmark calculations were performed for the HEAT dataset, 
\cite{HEAT1,HEAT2,HEAT3,HEAT4} 
which contains 31 atoms and molecules, and we compare our aug-cc-pVTZ results 
for the total, atomisation, and formation energies of these systems with
the complete-basis-set/full coupled-cluster extrapolated reference values from
Ref.~\onlinecite{HEAT1}, and with the xTC and F12 results from 
Ref.~\onlinecite{christlmaierXTC2023}.
The original orbitals were optimised at the HF and restricted open-shell HF level,
and the xTC integrals were evaluated using Hartree-Fock 1-RDMs.
Unless stated otherwise, all-electron calculations were performed 
and spin-resolved 1-RDMs were used for open-shell systems.

The cost of the biorthogonal orbital optimisation is negligible compared to the
cost of the xTC integral evaluation, and we have not encountered any convergence issues
in our test calculations.
\section{Results and Discussion}
\subsection{Total energies}
\begin{figure*}
\centering
  \includegraphics[height=10cm]{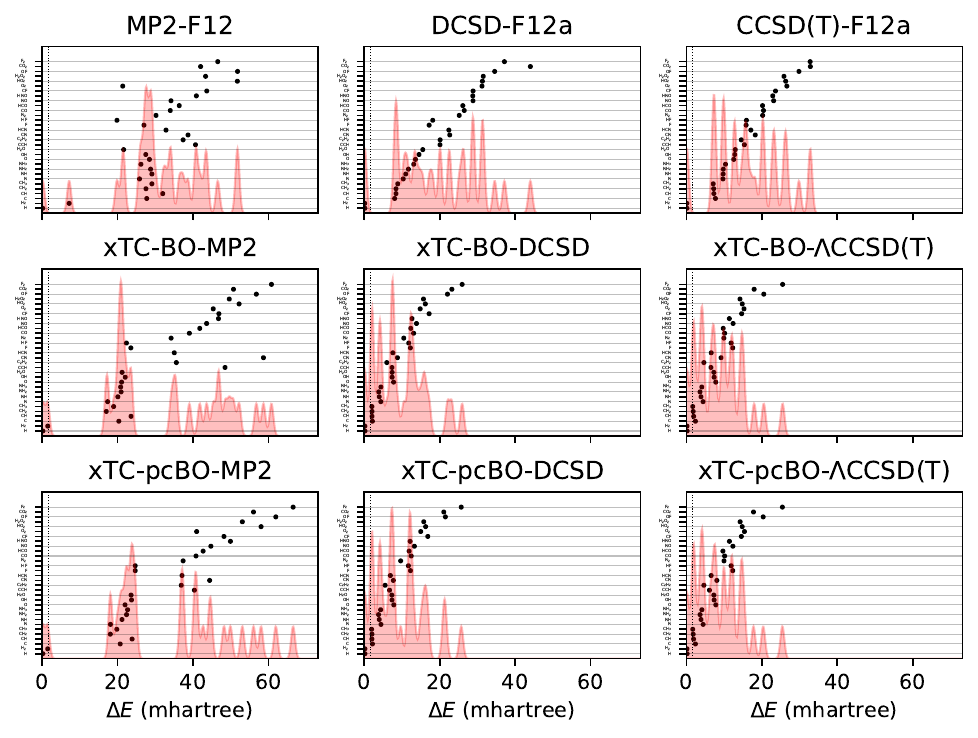}
  \caption{Errors in total energies of the atoms and molecules from the HEAT dataset,
  calculated using aug-cc-pVTZ basis set.
  The errors are calculated with respect to the extrapolated FCI/CBS limit from Ref.~\onlinecite{HEAT1}.
  BO and pcBO denote methods based on the biorthogonal orbital optimisation and 
  biorthogonal pseudo-canonical orbital transformation, respectively.
  Dotted lines indicate chemical accuracy (1 kcal/mol).
  The shaded area corresponds to the sum of Gaussians centered at each data point,
  with the width chosen such that for equally spaced points the Gaussians would be to 95\% 
  contained within their respective intervals.
  }
  \label{fig:totalenergy}
\end{figure*}

The total energy errors of the atoms and molecules from the HEAT dataset 
are shown in \fig{fig:totalenergy} and the corresponding statistics in terms of 
mean-signed deviation (MSD), standard deviation (STD) and maximal deviation (MaxD)
are summarised in \tab{tab:totalenergy}.
The biorthogonal orbital optimisation only slightly affects the accuracy of the xTC-DCSD method.
Note that the xTC-DCSD on top of pseudo-canonicalised orbitals (xTC-pcBO-DCSD) yields exactly 
the same results as the original xTC-DCSD method, 
and therefore the xTC-DCSD results are not shown in the figure.
In agreement with our previous xTC-CCSD and xTC-DCSD experience, 
the xTC-CCSD(T) total energies for both versions 
of biorthogonal orbital rotations are more accurate than CCSD(T)-F12 ones.

\begin{table}[h]
\small
\caption{\ Statistical measures of errors in total energies
(aug-cc-pVTZ basis)
with respect to HEAT estimates, in millihartree.}
\label{tab:totalenergy}
\begin{tabular*}{0.48\textwidth}{@{\extracolsep{\fill}}lrrr}
\hline
Method              & MSD & STD & MaxD \\
\hline
CCSD(T)-F12         & 16.4& 8.7 & 32.9 \\
DCSD-F12            & 20.0& 10.9& 44.0 \\
MP2-F12             & 31.7& 11.3& 51.8 \\
xTC-BO-$\Lambda$CCSD(T)&8.9& 6.2& 25.5\\
xTC-pcBO-$\Lambda$CCSD(T)&8.8&6.2&25.4\\
xTC-BO-DCSD         & 9.7 & 6.8 & 25.9\\
xTC-DCSD            & 9.4 & 6.6 & 25.7\\
xTC-BO-MP2          & 32.8& 16.4& 60.8\\
xTC-pcBO-MP2        & 33.7& 16.7& 66.5\\
\hline
\end{tabular*}
\end{table}

Surprisingly, the transcorrelated MP2 total energies (both, xTC-BO-MP2 and xTC-pcBO-MP2) 
for some systems, \textit{e.g.}, F$_2$, CO$_2$ or OF, turn out to be noticeably 
less accurate than MP2-F12. This hints to a potential limitation of the Jastrow optimization 
based on the minimisation of the variance of the reference energy, \eq{eq:variance2},
especially for perturbative methods:
the expression, which looks very similar to the xTC-MP2 energy expression, \eq{eq:xtc-mp2}, 
does not include the usual orbital-energy denominators, and as the result the integral 
contributions are weighted uniformly and not according to the importance in the correlation.
Unfortunately, inclusion of the orbital energies into the VMC framework is not feasible,
however, it is possible to optimize the Jastrow factors by using \eq{eq:xtc-mp2} directly,
and we are currently investigating this approach in our laboratory. 

High accuracy of the absolute energies does not necessarily translate into high accuracy of 
relative energies, which is much more important for applications. 
In the next sections we investigate the accuracy of transcorrelated methods based on 
biorthogonally optimised orbitals for computation of atomisation and formation energies.

\subsection{Atomisation energies}

\begin{figure*}
\centering
  \includegraphics[height=10cm]{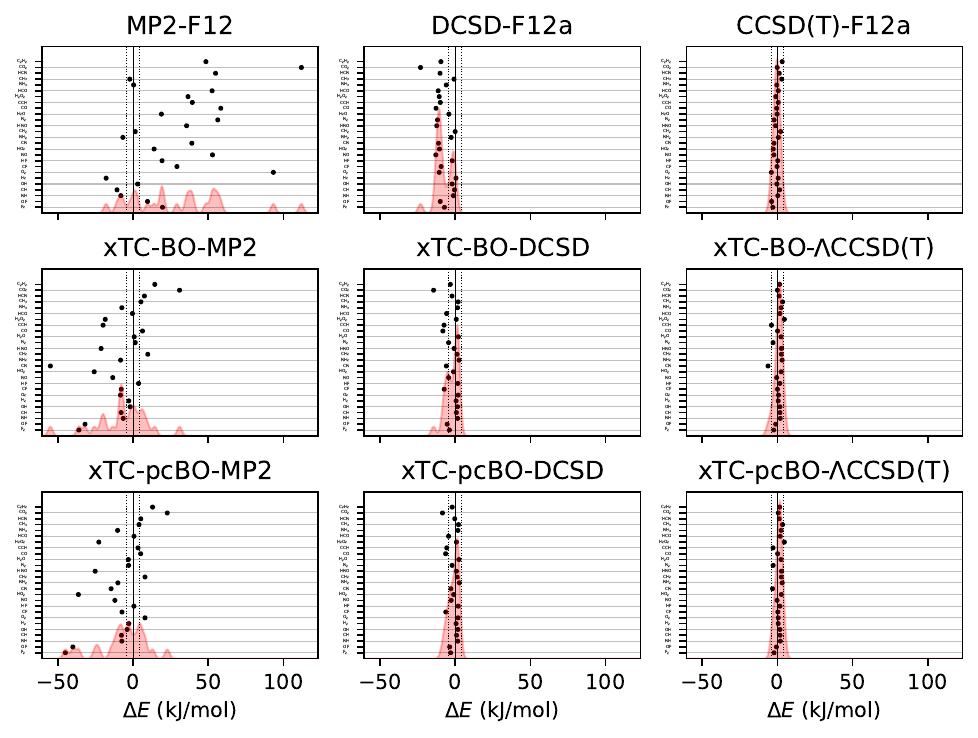}
  \caption{Errors in atomisation energies of molecules from the HEAT dataset,
  calculated using aug-cc-pVTZ basis set.
  The errors are calculated with respect to the extrapolated FCI/CBS limit from Ref.~\onlinecite{HEAT1}.
  BO and pcBO denote methods based on the biorthogonal orbital optimisation and 
  biorthogonal pseudo-canonical orbital transformation, respectively.
  Dotted lines indicate chemical accuracy (1 kcal/mol).
  The shaded area corresponds to the sum of Gaussians centered at each data point,
  with the width chosen such that for equally spaced points the Gaussians would be to 95\% 
  contained within their respective intervals.
  }
  \label{fig:atomisation}
\end{figure*}

The errors in atomisation energies of the molecules from the HEAT dataset 
are shown in \fig{fig:atomisation}, and the corresponding statistics in terms of 
mean-absolute deviation (MAD), root-mean squared deviation (RMSD) and maximal deviation (MaxD)
are summarised in \tab{tab:atomisation}.

\begin{table}[h]
\small
\caption{\ Statistical measures of errors in atomisation energies 
(aug-cc-pVTZ basis)
with respect to HEAT estimates,
 in kJ/mol.}
\label{tab:atomisation}
\begin{tabular*}{0.48\textwidth}{@{\extracolsep{\fill}}lrrr}
\hline
Method                  & MAD & RMSD& MaxD \\
\hline
CCSD(T)-F12             & 1.47& 1.90& -4.01\\
DCSD-F12                & 7.82& 9.50& -23.13\\
MP2-F12                 &32.30&42.59& 111.92\\
xTC-BO-$\Lambda$CCSD(T) & 2.07& 2.51& -6.20\\
xTC-pcBO-$\Lambda$CCSD(T)&1.94& 2.24& 4.63\\
xTC-BO-DCSD             & 3.56& 4.74& -14.42\\
xTC-DCSD                & 2.69& 3.35& -8.49\\
xTC-BO-MP2              &13.61&18.78& -55.03\\
xTC-pcBO-MP2            &12.40&17.31& -45.06\\
\hline
\end{tabular*}
\end{table}

The biorthogonal orbital optimisation noticeably worsens the accuracy of the xTC-DCSD method,
with RMSD increasing by 40\%.
As discussed in \sect{sec:xtc}, the orbital optimisation changes the reference determinant
and therefore the Jastrow factor is no longer optimal for the reference determinant of the
coupled-cluster calculations, and the accuracy of the transcorrelated results can deteriorate.

The xTC-CCSD(T) atomisation energies are more accurate than the xTC-DCSD ones,
and approach the accuracy of the CCSD(T)-F12 results. However, also in this case, the xTC-CCSD(T) 
method based on the biorthogonal orbital optimisation is less accurate than the 
one based on the pseudo-canonicalisation of the orbitals, although the difference is 
less pronounced than in the case of the xTC-DCSD method.

Interestingly, the xTC-MP2 atomisation energies are much more accurate than the ones obtained
from MP2-F12. This is in contrast to the total energies, and suggests that the Jastrow factor
optimisation based on the minimisation of the variance of the reference energy, \eq{eq:variance2},
yields balanced Jastrow factors, even if they are not minimising the xTC-MP2 correlation energy
contribution. 
Again, the xTC-MP2 results based on the biorthogonal orbital optimisation are less accurate
than the ones based on the pseudo-canonicalisation of the orbitals.

\subsection{Formation energies}

\begin{figure*}
\centering
  \includegraphics[height=10cm]{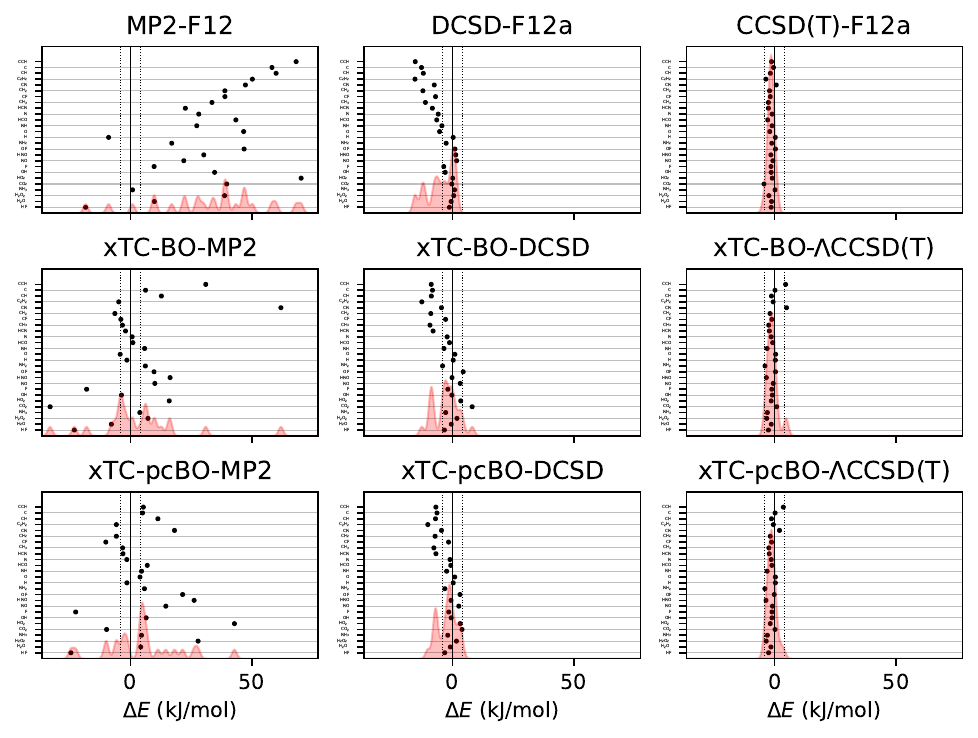}
  \caption{Errors in formation energies of molecules from the HEAT dataset,
  calculated using aug-cc-pVTZ basis set.
  The errors are calculated with respect to the extrapolated FCI/CBS limit from Ref.~\onlinecite{HEAT1}.
  BO and pcBO denote methods based on the biorthogonal orbital optimisation and 
  biorthogonal pseudo-canonical orbital transformation, respectively.
  Dotted lines indicate chemical accuracy (1 kcal/mol).
  The shaded area corresponds to the sum of Gaussians centered at each data point,
  with the width chosen such that for equally spaced points the Gaussians would be to 95\% 
  contained within their respective intervals.
  }
  \label{fig:reaction}
\end{figure*}

Formation energies of the molecules from the HEAT dataset 
(see Table I from Ref.~\onlinecite{christlmaierXTC2023}) have been calculated using the
transcorrelated methods for biorthogonally optimised orbitals,
and the errors with respect to the extrapolated full coupled cluster results at the 
complete basis set limit from Ref.~\onlinecite{HEAT1} are shown in \fig{fig:reaction}. 
The statistics of the errors are summarised in \tab{tab:reaction}.

\begin{table}[h]
\small
\caption{\ Statistical measures of errors in formation energies
(aug-cc-pVTZ basis)
with respect to HEAT estimates, in kJ/mol.}
\label{tab:reaction}
\begin{tabular*}{0.48\textwidth}{@{\extracolsep{\fill}}lrrr}
\hline
Method                  & MAD & RMSD& MaxD \\
\hline
CCSD(T)-F12             & 1.60& 1.89& -4.37\\
DCSD-F12                & 5.42& 7.32& -15.47\\
MP2-F12                 &35.05&39.43& 70.33\\
xTC-BO-$\Lambda$CCSD(T) & 1.86& 2.29& 4.89\\
xTC-pcBO-$\Lambda$CCSD(T)&1.73& 2.08& -4.02\\
xTC-BO-DCSD             & 4.40& 5.58& -12.64\\
xTC-DCSD                & 3.47& 4.40& -10.17\\
xTC-BO-MP2              &11.57&17.55&61.99\\
xTC-pcBO-MP2            &11.43&15.30&42.85\\
\hline
\end{tabular*}
\end{table}

The results for the formation energies lead to similar conclusions as for the atomisation energies.
The xTC-DCSD results on top of the biorthogonally optimised orbitals are less accurate than
the original xTC-DCSD results, 
and the xTC-$\Lambda$CCSD(T) results are more accurate than the xTC-DCSD ones. 
As before, the sensitivity of the xTC-CCSD(T) results to the orbital optimisation is less 
pronounced compared to the xTC-DCSD results.
The xTC-DCSD results are considerably more accurate than DCSD-F12, and the xTC-CCSD(T) results
are close in the accuracy to the CCSD(T)-F12 results.
The xTC-MP2 formation energies are much more accurate than the MP2-F12 ones, which 
again demonstrates the balanced description of the correlation by the Jastrow factors. 

\subsection{Effect of the xTC approximation}
In order to investigate the accuracy of the xTC approximation, we have performed
calculations of the total, atomisation, and formation energies using the xTC methods
with the spin-averaged 1-RDMs, which yields spin-independent xTC integrals.
In our previous calculations \cite{christlmaierXTC2023}, we have found that xTC-DCSD based on
the spin-independent xTC integrals are as accurate as the spin-dependent ones.

\begin{table*}
\small
\caption{\ Statistical measures of errors in total, atomisation, and formation energies 
(aug-cc-pVTZ basis)
with respect to HEAT estimates for xTC approximation using spin-averaged 1-RDMs.}
\label{tab:spinavg}
\begin{tabular*}{\textwidth}{@{\extracolsep{\fill}}lrrrrrrrrr}
\hline
Method                  & \multicolumn{3}{c}{Total energy, mE$_h$} & \multicolumn{3}{c}{Atomisation energy, kJ/mol} &
\multicolumn{3}{c}{Formation energy, kJ/mol}\\
                        & MSD & STD& MaxD & MAD & RMSD& MaxD & MAD & RMSD& MaxD \\
\hline
xTC-BO-$\Lambda$CCSD(T) & 9.1& 6.1& 25.5& 3.62& 4.12& 8.79 & 2.23& 3.07& 9.27\\
xTC-pcBO-$\Lambda$CCSD(T)&9.0& 6.1& 25.4& 3.56& 4.15& 8.78 & 2.07& 2.72& 8.33\\
xTC-BO-DCSD             & 9.9& 6.7& 25.9& 2.95& 3.57&-8.78 & 3.26& 4.18&10.29\\
xTC-DCSD                & 9.6& 6.5& 25.7& 2.67& 3.09& 6.14 & 2.52& 3.02& 6.14\\
xTC-BO-MP2              &33.0&16.3& 60.8&13.21&18.04&-52.62&11.63&18.21&64.63\\
xTC-pcBO-MP2            &34.0&16.6& 66.5&12.03&16.47&-42.54&11.63&15.68&43.61\\
\hline
\end{tabular*}
\end{table*}

The statistics of errors in the total, atomisation, and formation energies of the atoms and 
molecules from the HEAT dataset are summarised in \tab{tab:spinavg}. 
The total energies of all methods are hardly affected by the different choice of the 1-RDMs 
in the xTC approximation.
In agreement with our previous results,
the relative xTC-DCSD energies based on xTC integrals calculated using the spin-averaged 1-RDMs are
more accurate than the xTC-DCSD energies based on xTC integrals with the spin-resolved 
density matrices, and the biorthogonal orbital optimisation reduces the accuracy of the xTC-DCSD.

On the other hand, the xTC-CCSD(T) atomisation and formation energies are clearly less accurate 
when the spin-averaged-1-RDM based xTC integrals are used, 
and the accuracy of xTC-MP2 is comparable to the one based on the spin-resolved 1-RDMs.
This suggests that the accuracy of the xTC approximation starts to become one of the limiting 
factors for the xTC-CCSD(T) method, 
and the choice of the 1-RDMs in the xTC approximation is important
to obtain accurate results. The accuracy of the xTC approximation can be improved by using
perturbative corrections to account for the missing explicit three-body terms in the xTC Hamiltonian,
however, this would require a substantial increase in the computational cost of the xTC integrals.

\subsection{Frozen-core calculations}

One of the advantages of the transcorrelated methods based on optimised Jastrow factors 
is the possibility to perform calculations 
with frozen-core approximation with minimal loss of accuracy, as has been demonstrated 
for the xTC-DCSD method in Ref.~\onlinecite{christlmaierXTC2023}.
On the other hand, Ammar \textit{et al.} \cite{ammarTranscorrelated2023} have shown that the
accuracy of the frozen-core approximation in transcorrelated methods can be improved 
by using the biorthogonal orbital optimisation, and without the orbital optimisation the 
frozen-core approximation leads to large errors in the transcorrelated methods with 
atomic Jastrow factors.
Thus, to assess the effect of the biorthogonal orbital optimisation on the accuracy of the
frozen-core calculations, we have performed such calculations of the total, atomisation, and
formation energies of the atoms and molecules from the HEAT dataset using the xTC methods.
The core electrons were frozen after the orbital optimisation, and we compare the results
with the ones without the orbital optimisation and with the F12 results.
The statistics of the errors are summarised in \tab{tab:frozen}.

\begin{table*}
\small
\caption{\ Statistical measures of errors in total, atomisation, and formation energies 
(aug-cc-pVTZ basis)
with respect to HEAT estimates for frozen-core calculations.} 
\label{tab:frozen}
\begin{tabular*}{\textwidth}{@{\extracolsep{\fill}}lrrrrrrrrr}
\hline
Method                  & \multicolumn{3}{c}{Total energy, mE$_h$} & \multicolumn{3}{c}{Atomisation energy, kJ/mol} &
\multicolumn{3}{c}{Formation energy, kJ/mol}\\
                        & MSD & STD& MaxD & MAD & RMSD& MaxD & MAD & RMSD& MaxD \\
\hline
CCSD(T)-F12             & 95.3& 44.4&194.0& 5.47& 6.53&-12.78& 1.91& 2.57& -7.05\\
DCSD-F12                & 98.7& 46.7&204.9&12.79&15.16&-35.50& 3.68& 4.80&-12.02\\
MP2-F12                 &110.7& 45.9&203.0&29.77&38.87&102.26&34.85&39.03& 70.84\\
xTC-BO-$\Lambda$CCSD(T) & 12.6& 7.1 & 28.9& 2.88& 4.01&-10.81& 2.10& 2.38& 3.88\\
xTC-pcBO-$\Lambda$CCSD(T)&10.8& 6.2 & 25.5& 2.19& 2.88& -7.40& 2.03& 2.21& -3.50\\
xTC-BO-DCSD             & 13.4& 7.9 & 31.1& 5.68& 7.63&-20.84& 5.00& 6.39&-15.06\\
xTC-pcBO-DCSD           & 11.3& 6.7 & 25.8& 3.91& 5.20&-12.43& 3.75& 4.87&-12.17\\
xTC-DCSD                & 11.3& 6.7 & 25.8& 3.79& 5.06&-12.13& 3.68& 4.80&-12.02\\
xTC-BO-MP2              & 36.1& 17.6& 64.0&14.14&19.74&-58.42&11.69&17.59& 61.73\\
xTC-pcBO-MP2            & 34.9& 16.8& 65.5&12.38&17.56&-45.83&11.55&15.46& 42.91\\
\hline
\end{tabular*}
\end{table*}

For the frozen core calculations xTC-pcBO-DCSD and xTC-DCSD results differ from each other,
but only slightly, which suggests that the core orbitals and the remaining occupied orbitals
are not strongly mixed in the pseudo-canonicalisation of the orbitals.
Comparing the xTC-DCSD results among themselves, the biorthogonal orbital optimisation
does not improve the accuracy of the frozen-core approximation in our calculations; on 
the contrary, the difference in the accuracy of the all-electron xTC-DCSD and xTC-BO-DCSD 
results is smaller, than the difference in the accuracy of the frozen-core xTC-DCSD and 
xTC-BO-DCSD results. It means that also in this case the biorthogonal orbital optimisation
does not help to improve the accuracy of the transcorrelated calculations.
Again, we attribute this to the fact that our Jastrow factors are optimised for the
molecules according to \eq{eq:variance2}, and the orbital optimisation changes the reference
determinant.

Atomisation energies from the frozen-core xTC-MP2 method with pseudo-canonical orbitals 
approach the accuracy of the frozen-core DCSD-F12, but the formation energies are less accurate.
Nevertheless, the frozen-core xTC-MP2 results are much more accurate than the frozen-core 
(and all-electron) MP2-F12 results.

The xTC-CCSD(T) results based on the pseudo-canonicalised orbitals are the most accurate 
ones among all methods employed in these frozen-core calculations. Compared to the all-electron
xTC-CCSD(T) results, the accuracy of the frozen-core xTC-CCSD(T) results is slightly worse,
but still better than the accuracy of the frozen-core CCSD(T)-F12 results.

\section{Conclusions}

In this work, we have investigated the effect of the biorthogonal orbital optimisation on the
accuracy of the transcorrelated methods based on the xTC approximation and Jastrow factors
optimised for the reference determinant through the minimisation of the variance of the
reference energy.
Additionally, we have investigated the accuracy of the xTC approximation in the combination
with M{\o}ller-Plesset perturbation theory based methods, MP2 and CCSD(T). 
For CCSD(T) on the xTC Hamiltonian, we have employed the $\Lambda$CCSD(T) method, 
which is very similar to the standard CCSD(T) method, but does not rely on the hermiticity
of the Hamiltonian.

In all our benchmark calculations, the biorthogonal orbital optimisation has not improved 
the accuracy of the xTC based coupled-cluster methods, 
and in most cases it has even worsened the accuracy of the transcorrelated results. 
This can be attributed to the fact that the Jastrow factors are optimised for the reference
determinant, to minimise the residual correlation with respect to this determinant, 
and the orbital optimisation changes the reference, and therefore the
Jastrow factors are no longer optimal for the reference determinant of the coupled-cluster
calculations.

As an alternative to the biorthogonal orbital optimisation, we have investigated the
pseudo-canonicalisation of the orbitals, and found that the xTC-CCSD(T) results based on the
pseudo-canonicalised orbitals are more accurate than the ones based on the biorthogonally
optimised orbitals, and are on par with the CCSD(T)-F12 results.
Obviously, the higher excitations are included into the coupled cluster method, the
less sensitive the results are to the orbital optimisation, and the xTC-CCSD(T) results 
based on the pseudo-canonicalised orbitals are much closer in the accuracy to the 
orbital-optimised xTC-CCSD(T) results, than in the case of the xTC-DCSD results.

As in our previous work, \cite{christlmaierXTC2023} the frozen-core xTC results are 
very accurate for all methods, and the xTC-$\Lambda$CCSD(T) results based on the 
pseudo-canonicalised orbitals are the most accurate ones among all methods employed in
this work. The biorthogonal orbital optimisation does not improve the accuracy of the
frozen-core calculations. 
This is in contrast to the results of Ammar \textit{et al.} \cite{ammarTranscorrelated2023},
who have used atomic Jastrow factors, and found that the biorthogonal orbital optimisation
greatly improves the accuracy of the frozen-core calculations.

The xTC-MP2 results are generally much more accurate than the MP2-F12 results,
however, total energies of some molecules are less accurate than the MP2-F12 ones.
This suggests that the Jastrow factor optimisation based on the minimisation of the variance
of the reference energy, \eq{eq:variance2}, can be improved by including the orbital energies
as the weights for the integral contributions, which would lead to minimisation of the 
xTC-MP2 correlation energy, \eq{eq:xtc-mp2}, and we are currently investigating this approach.

The somewhat sobering results of the xTC approximation in the combination with the
CCSD(T) method compared to the all-electron CCSD(T)-F12 results, 
suggest that there is still room for improvement of the xTC approximation and the Jastrow factor
optimisation.
The accuracy of the xTC approximation is one of the limiting factors for the xTC-CCSD(T) method,
and the choice of the 1-RDMs in the xTC approximation is important to obtain accurate results.
Besides, the stochastic errors in the VMC calculations for the Jastrow optimisation lead to
non-systematic errors in the final energies and the worse error cancellation 
in the relative energies.

The new implemented xTC-$\Lambda$CCSD(T) method will be useful to investigate the accuracy 
of the alternative ways of optimising the Jastrow factors and improving the xTC approximation.
The biorthogonal orbital optimisation can become important in the cases where the Jastrow factors 
are not optimal for the reference determinant of the subsequent coupled cluster methods,
\textit{e.g.}, for transferable Jastrow factors which can benefit more from error cancellation
in the relative energies, and we are currently working on such an approach in our laboratory.


\section*{Conflicts of interest}
There are no conflicts to declare.

\section*{Acknowledgements}
Funded by the Deutsche Forschungsgemeinschaft (DFG, German Research Foundation) -- 455145945. 
Financial support from the Max-Planck Society is gratefully acknowledged.


\balance


\bibliography{boxtc,molpro} 

\providecommand*{\mcitethebibliography}{\thebibliography}
\csname @ifundefined\endcsname{endmcitethebibliography}
{\let\endmcitethebibliography\endthebibliography}{}
\begin{mcitethebibliography}{118}
\providecommand*{\natexlab}[1]{#1}
\providecommand*{\mciteSetBstSublistMode}[1]{}
\providecommand*{\mciteSetBstMaxWidthForm}[2]{}
\providecommand*{\mciteBstWouldAddEndPuncttrue}
  {\def\EndOfBibitem{\unskip.}}
\providecommand*{\mciteBstWouldAddEndPunctfalse}
  {\let\EndOfBibitem\relax}
\providecommand*{\mciteSetBstMidEndSepPunct}[3]{}
\providecommand*{\mciteSetBstSublistLabelBeginEnd}[3]{}
\providecommand*{\EndOfBibitem}{}
\mciteSetBstSublistMode{f}
\mciteSetBstMaxWidthForm{subitem}
{(\emph{\alph{mcitesubitemcount}})}
\mciteSetBstSublistLabelBeginEnd{\mcitemaxwidthsubitemform\space}
{\relax}{\relax}

\bibitem[\v{C}\'{\i}\v{z}ek(1966)]{cizek:66}
J.~\v{C}\'{\i}\v{z}ek, \emph{J. Chem. Phys.}, 1966, \textbf{45},
  4256--4266\relax
\mciteBstWouldAddEndPuncttrue
\mciteSetBstMidEndSepPunct{\mcitedefaultmidpunct}
{\mcitedefaultendpunct}{\mcitedefaultseppunct}\relax
\EndOfBibitem
\bibitem[Hampel and Werner(1996)]{Hampel:96}
C.~Hampel and H.-J. Werner, \emph{J. Chem. Phys.}, 1996, \textbf{104},
  6286--6297\relax
\mciteBstWouldAddEndPuncttrue
\mciteSetBstMidEndSepPunct{\mcitedefaultmidpunct}
{\mcitedefaultendpunct}{\mcitedefaultseppunct}\relax
\EndOfBibitem
\bibitem[Sch{\"u}tz(2000)]{Schuetz:00b}
M.~Sch{\"u}tz, \emph{J. Chem. Phys.}, 2000, \textbf{113}, 9986--10001\relax
\mciteBstWouldAddEndPuncttrue
\mciteSetBstMidEndSepPunct{\mcitedefaultmidpunct}
{\mcitedefaultendpunct}{\mcitedefaultseppunct}\relax
\EndOfBibitem
\bibitem[Werner and Sch{\"u}tz(2011)]{Werner:2011a}
H.-J. Werner and M.~Sch{\"u}tz, \emph{J. Chem. Phys.}, 2011, \textbf{135},
  144116\relax
\mciteBstWouldAddEndPuncttrue
\mciteSetBstMidEndSepPunct{\mcitedefaultmidpunct}
{\mcitedefaultendpunct}{\mcitedefaultseppunct}\relax
\EndOfBibitem
\bibitem[Riplinger \emph{et~al.}(2016)Riplinger, Pinski, Becker, Valeev, and
  Neese]{Riplinger:2016}
C.~Riplinger, P.~Pinski, U.~Becker, E.~F. Valeev and F.~Neese, \emph{J. Chem.
  Phys.}, 2016, \textbf{144}, 024109\relax
\mciteBstWouldAddEndPuncttrue
\mciteSetBstMidEndSepPunct{\mcitedefaultmidpunct}
{\mcitedefaultendpunct}{\mcitedefaultseppunct}\relax
\EndOfBibitem
\bibitem[Schmitz and H{\"a}ttig(2016)]{Schmitz:2016}
G.~Schmitz and C.~H{\"a}ttig, \emph{J. Chem. Phys.}, 2016, \textbf{145},
  234107\relax
\mciteBstWouldAddEndPuncttrue
\mciteSetBstMidEndSepPunct{\mcitedefaultmidpunct}
{\mcitedefaultendpunct}{\mcitedefaultseppunct}\relax
\EndOfBibitem
\bibitem[Schwilk \emph{et~al.}(2015)Schwilk, Usvyat, and Werner]{Schwilk:2015}
M.~Schwilk, D.~Usvyat and H.-J. Werner, \emph{J. Chem. Phys.}, 2015,
  \textbf{142}, 121102\relax
\mciteBstWouldAddEndPuncttrue
\mciteSetBstMidEndSepPunct{\mcitedefaultmidpunct}
{\mcitedefaultendpunct}{\mcitedefaultseppunct}\relax
\EndOfBibitem
\bibitem[Schwilk \emph{et~al.}(2017)Schwilk, Ma, K\"{o}ppl, and
  Werner]{Schwilk:2017}
M.~Schwilk, Q.~Ma, C.~K\"{o}ppl and H.-J. Werner, \emph{J. Chem. Theory
  Comput.}, 2017, \textbf{13}, 3650--3675\relax
\mciteBstWouldAddEndPuncttrue
\mciteSetBstMidEndSepPunct{\mcitedefaultmidpunct}
{\mcitedefaultendpunct}{\mcitedefaultseppunct}\relax
\EndOfBibitem
\bibitem[Ma and Werner(2017)]{Ma:2017a}
Q.~Ma and H.-J. Werner, \emph{J. Chem. Theory Comput.}, 2017, \textbf{14},
  198--215\relax
\mciteBstWouldAddEndPuncttrue
\mciteSetBstMidEndSepPunct{\mcitedefaultmidpunct}
{\mcitedefaultendpunct}{\mcitedefaultseppunct}\relax
\EndOfBibitem
\bibitem[Meyer(1971)]{Meyer:71}
W.~Meyer, \emph{Int. J. Quantum Chem. Symp.}, 1971, \textbf{5}, 341\relax
\mciteBstWouldAddEndPuncttrue
\mciteSetBstMidEndSepPunct{\mcitedefaultmidpunct}
{\mcitedefaultendpunct}{\mcitedefaultseppunct}\relax
\EndOfBibitem
\bibitem[Paldus \emph{et~al.}(1984)Paldus, \v{C}\'{\i}\v{z}ek, and
  Takahashi]{paldus_approximate_1984}
J.~Paldus, J.~\v{C}\'{\i}\v{z}ek and M.~Takahashi, \emph{Phys. Rev. A}, 1984,
  \textbf{30}, 2193\relax
\mciteBstWouldAddEndPuncttrue
\mciteSetBstMidEndSepPunct{\mcitedefaultmidpunct}
{\mcitedefaultendpunct}{\mcitedefaultseppunct}\relax
\EndOfBibitem
\bibitem[Piecuch and Paldus(1991)]{piecuch_solution_1991}
P.~Piecuch and J.~Paldus, \emph{Int. J. Quantum Chem.}, 1991, \textbf{40},
  9--34\relax
\mciteBstWouldAddEndPuncttrue
\mciteSetBstMidEndSepPunct{\mcitedefaultmidpunct}
{\mcitedefaultendpunct}{\mcitedefaultseppunct}\relax
\EndOfBibitem
\bibitem[Piecuch \emph{et~al.}(1996)Piecuch, Tobola, and
  Paldus]{piecuch_approximate_1996}
P.~Piecuch, R.~Tobola and J.~Paldus, \emph{Phys. Rev. A}, 1996, \textbf{54},
  1210--1241\relax
\mciteBstWouldAddEndPuncttrue
\mciteSetBstMidEndSepPunct{\mcitedefaultmidpunct}
{\mcitedefaultendpunct}{\mcitedefaultseppunct}\relax
\EndOfBibitem
\bibitem[Kowalski and Piecuch(2000)]{kowalski_method_2000}
K.~Kowalski and P.~Piecuch, \emph{J. Chem. Phys.}, 2000, \textbf{113}, 18\relax
\mciteBstWouldAddEndPuncttrue
\mciteSetBstMidEndSepPunct{\mcitedefaultmidpunct}
{\mcitedefaultendpunct}{\mcitedefaultseppunct}\relax
\EndOfBibitem
\bibitem[Bartlett and Musia\l(2006)]{bartlett_addition_2006}
R.~J. Bartlett and M.~Musia\l, \emph{J. Chem. Phys.}, 2006, \textbf{125},
  204105\relax
\mciteBstWouldAddEndPuncttrue
\mciteSetBstMidEndSepPunct{\mcitedefaultmidpunct}
{\mcitedefaultendpunct}{\mcitedefaultseppunct}\relax
\EndOfBibitem
\bibitem[Nooijen and Le~Roy(2006)]{nooijen_orbital_2006}
M.~Nooijen and R.~J. Le~Roy, \emph{J. Mol. Struc.}, 2006, \textbf{768},
  25--43\relax
\mciteBstWouldAddEndPuncttrue
\mciteSetBstMidEndSepPunct{\mcitedefaultmidpunct}
{\mcitedefaultendpunct}{\mcitedefaultseppunct}\relax
\EndOfBibitem
\bibitem[Neese \emph{et~al.}(2009)Neese, Wennmohs, and Hansen]{Neese:09}
F.~Neese, F.~Wennmohs and A.~Hansen, \emph{J. Chem. Phys.}, 2009, \textbf{130},
  114108\relax
\mciteBstWouldAddEndPuncttrue
\mciteSetBstMidEndSepPunct{\mcitedefaultmidpunct}
{\mcitedefaultendpunct}{\mcitedefaultseppunct}\relax
\EndOfBibitem
\bibitem[Huntington and Nooijen(2010)]{huntington_pccsd:_2010}
L.~M.~J. Huntington and M.~Nooijen, \emph{J. Chem. Phys.}, 2010, \textbf{133},
  184109\relax
\mciteBstWouldAddEndPuncttrue
\mciteSetBstMidEndSepPunct{\mcitedefaultmidpunct}
{\mcitedefaultendpunct}{\mcitedefaultseppunct}\relax
\EndOfBibitem
\bibitem[Robinson and Knowles(2011)]{robinson_approximate_2011}
J.~B. Robinson and P.~J. Knowles, \emph{J. Chem. Phys.}, 2011, \textbf{135},
  044113\relax
\mciteBstWouldAddEndPuncttrue
\mciteSetBstMidEndSepPunct{\mcitedefaultmidpunct}
{\mcitedefaultendpunct}{\mcitedefaultseppunct}\relax
\EndOfBibitem
\bibitem[Huntington \emph{et~al.}(2012)Huntington, Hansen, Neese, and
  Nooijen]{huntington_accurate_2012}
L.~M.~J. Huntington, A.~Hansen, F.~Neese and M.~Nooijen, \emph{J. Chem. Phys.},
  2012, \textbf{136}, 064101\relax
\mciteBstWouldAddEndPuncttrue
\mciteSetBstMidEndSepPunct{\mcitedefaultmidpunct}
{\mcitedefaultendpunct}{\mcitedefaultseppunct}\relax
\EndOfBibitem
\bibitem[Paldus(2017)]{paldusExternally2017}
J.~Paldus, \emph{J. Mat. Chem.}, 2017, \textbf{55}, 477--502\relax
\mciteBstWouldAddEndPuncttrue
\mciteSetBstMidEndSepPunct{\mcitedefaultmidpunct}
{\mcitedefaultendpunct}{\mcitedefaultseppunct}\relax
\EndOfBibitem
\bibitem[Black and Knowles(2018)]{black_statistical_2018}
J.~A. Black and P.~J. Knowles, \emph{Mol. Phys.}, 2018, \textbf{116},
  1421\relax
\mciteBstWouldAddEndPuncttrue
\mciteSetBstMidEndSepPunct{\mcitedefaultmidpunct}
{\mcitedefaultendpunct}{\mcitedefaultseppunct}\relax
\EndOfBibitem
\bibitem[Behnle and Fink(2019)]{behnleREMP2019}
S.~Behnle and R.~F. Fink, \emph{J. Chem. Phys.}, 2019, \textbf{150},
  124107\relax
\mciteBstWouldAddEndPuncttrue
\mciteSetBstMidEndSepPunct{\mcitedefaultmidpunct}
{\mcitedefaultendpunct}{\mcitedefaultseppunct}\relax
\EndOfBibitem
\bibitem[Behnle and Fink(2021)]{behnleOOREMP2021}
S.~Behnle and R.~F. Fink, \emph{J. Chem. Theory Comput.}, 2021, \textbf{17},
  3259\relax
\mciteBstWouldAddEndPuncttrue
\mciteSetBstMidEndSepPunct{\mcitedefaultmidpunct}
{\mcitedefaultendpunct}{\mcitedefaultseppunct}\relax
\EndOfBibitem
\bibitem[Kats and Manby(2013)]{kats_dc_2013}
D.~Kats and F.~R. Manby, \emph{J. Chem. Phys.}, 2013, \textbf{139},
  021102\relax
\mciteBstWouldAddEndPuncttrue
\mciteSetBstMidEndSepPunct{\mcitedefaultmidpunct}
{\mcitedefaultendpunct}{\mcitedefaultseppunct}\relax
\EndOfBibitem
\bibitem[Kats(2014)]{kats_dcsd_2014}
D.~Kats, \emph{J. Chem. Phys.}, 2014, \textbf{141}, 061101\relax
\mciteBstWouldAddEndPuncttrue
\mciteSetBstMidEndSepPunct{\mcitedefaultmidpunct}
{\mcitedefaultendpunct}{\mcitedefaultseppunct}\relax
\EndOfBibitem
\bibitem[Kats \emph{et~al.}(2015)Kats, Kreplin, Werner, and
  Manby]{kats_accurate_2015}
D.~Kats, D.~Kreplin, H.-J. Werner and F.~R. Manby, \emph{J. Chem. Phys.}, 2015,
  \textbf{142}, 064111\relax
\mciteBstWouldAddEndPuncttrue
\mciteSetBstMidEndSepPunct{\mcitedefaultmidpunct}
{\mcitedefaultendpunct}{\mcitedefaultseppunct}\relax
\EndOfBibitem
\bibitem[Kats and K{\"o}hn(2019)]{katsDistinguishable2019}
D.~Kats and A.~K{\"o}hn, \emph{J. Chem. Phys.}, 2019, \textbf{150},
  151101\relax
\mciteBstWouldAddEndPuncttrue
\mciteSetBstMidEndSepPunct{\mcitedefaultmidpunct}
{\mcitedefaultendpunct}{\mcitedefaultseppunct}\relax
\EndOfBibitem
\bibitem[Rishi and Valeev(2019)]{rishiCan2019}
V.~Rishi and E.~F. Valeev, \emph{J. Chem. Phys.}, 2019, \textbf{151},
  064102\relax
\mciteBstWouldAddEndPuncttrue
\mciteSetBstMidEndSepPunct{\mcitedefaultmidpunct}
{\mcitedefaultendpunct}{\mcitedefaultseppunct}\relax
\EndOfBibitem
\bibitem[Kats(2018)]{kats_improving_2018}
D.~Kats, \emph{Mol. Phys.}, 2018, \textbf{116}, 1435\relax
\mciteBstWouldAddEndPuncttrue
\mciteSetBstMidEndSepPunct{\mcitedefaultmidpunct}
{\mcitedefaultendpunct}{\mcitedefaultseppunct}\relax
\EndOfBibitem
\bibitem[Rishi \emph{et~al.}(2017)Rishi, Perera, Nooijen, and
  Bartlett]{rishi_excited_2017}
V.~Rishi, A.~Perera, M.~Nooijen and R.~J. Bartlett, \emph{J. Chem. Phys.},
  2017, \textbf{146}, 144104\relax
\mciteBstWouldAddEndPuncttrue
\mciteSetBstMidEndSepPunct{\mcitedefaultmidpunct}
{\mcitedefaultendpunct}{\mcitedefaultseppunct}\relax
\EndOfBibitem
\bibitem[Tsatsoulis \emph{et~al.}(2017)Tsatsoulis, Hummel, Usvyat, Sch{\"u}tz,
  Booth, Binnie, Gillan, Alf{\'e}, Michaelides, and
  Gr{\"u}neis]{tsatsoulis_comparison_2017}
T.~Tsatsoulis, F.~Hummel, D.~Usvyat, M.~Sch{\"u}tz, G.~H. Booth, S.~S. Binnie,
  M.~J. Gillan, D.~Alf{\'e}, A.~Michaelides and A.~Gr{\"u}neis, \emph{J. Chem.
  Phys.}, 2017, \textbf{146}, 204108\relax
\mciteBstWouldAddEndPuncttrue
\mciteSetBstMidEndSepPunct{\mcitedefaultmidpunct}
{\mcitedefaultendpunct}{\mcitedefaultseppunct}\relax
\EndOfBibitem
\bibitem[Li~Manni \emph{et~al.}(2019)Li~Manni, Kats, Tew, and
  Alavi]{limanniRole2019}
G.~Li~Manni, D.~Kats, D.~P. Tew and A.~Alavi, \emph{J. Chem. Theory Comput.},
  2019, \textbf{15}, 1492\relax
\mciteBstWouldAddEndPuncttrue
\mciteSetBstMidEndSepPunct{\mcitedefaultmidpunct}
{\mcitedefaultendpunct}{\mcitedefaultseppunct}\relax
\EndOfBibitem
\bibitem[Vitale \emph{et~al.}(2020)Vitale, Alavi, and
  Kats]{vitaleFCIQMCTailored2020}
E.~Vitale, A.~Alavi and D.~Kats, \emph{J. Chem. Theory Comput.}, 2020,
  \textbf{16}, 5621\relax
\mciteBstWouldAddEndPuncttrue
\mciteSetBstMidEndSepPunct{\mcitedefaultmidpunct}
{\mcitedefaultendpunct}{\mcitedefaultseppunct}\relax
\EndOfBibitem
\bibitem[Lin \emph{et~al.}(2020)Lin, Maschio, Kats, Usvyat, and
  Heine]{linFragmentBased2020}
H.-H. Lin, L.~Maschio, D.~Kats, D.~Usvyat and T.~Heine, \emph{J. Chem. Theory
  Comput.}, 2020, \textbf{16}, 7100\relax
\mciteBstWouldAddEndPuncttrue
\mciteSetBstMidEndSepPunct{\mcitedefaultmidpunct}
{\mcitedefaultendpunct}{\mcitedefaultseppunct}\relax
\EndOfBibitem
\bibitem[Schraivogel and Kats(2021)]{schraivogelAccuracy2021}
T.~Schraivogel and D.~Kats, \emph{J. Chem. Phys.}, 2021, \textbf{155},
  064101\relax
\mciteBstWouldAddEndPuncttrue
\mciteSetBstMidEndSepPunct{\mcitedefaultmidpunct}
{\mcitedefaultendpunct}{\mcitedefaultseppunct}\relax
\EndOfBibitem
\bibitem[Kutzelnigg(1985)]{k1985}
W.~Kutzelnigg, \emph{Theor. Chim. Acta}, 1985, \textbf{68}, 445\relax
\mciteBstWouldAddEndPuncttrue
\mciteSetBstMidEndSepPunct{\mcitedefaultmidpunct}
{\mcitedefaultendpunct}{\mcitedefaultseppunct}\relax
\EndOfBibitem
\bibitem[Kutzelnigg and Klopper(1991)]{kk1991i}
W.~Kutzelnigg and W.~Klopper, \emph{J. Chem. Phys.}, 1991, \textbf{94},
  1985--2001\relax
\mciteBstWouldAddEndPuncttrue
\mciteSetBstMidEndSepPunct{\mcitedefaultmidpunct}
{\mcitedefaultendpunct}{\mcitedefaultseppunct}\relax
\EndOfBibitem
\bibitem[Klopper and Samson(2002)]{ks2002}
W.~Klopper and C.~C.~M. Samson, \emph{J. Chem. Phys.}, 2002, \textbf{116},
  6397--6410\relax
\mciteBstWouldAddEndPuncttrue
\mciteSetBstMidEndSepPunct{\mcitedefaultmidpunct}
{\mcitedefaultendpunct}{\mcitedefaultseppunct}\relax
\EndOfBibitem
\bibitem[Manby(2003)]{m2003}
F.~R. Manby, \emph{J. Chem. Phys.}, 2003, \textbf{119}, 4607--4613\relax
\mciteBstWouldAddEndPuncttrue
\mciteSetBstMidEndSepPunct{\mcitedefaultmidpunct}
{\mcitedefaultendpunct}{\mcitedefaultseppunct}\relax
\EndOfBibitem
\bibitem[Ten-no(2004)]{tenno2004}
S.~Ten-no, \emph{Chem. Phys. Lett.}, 2004, \textbf{398}, 56--61\relax
\mciteBstWouldAddEndPuncttrue
\mciteSetBstMidEndSepPunct{\mcitedefaultmidpunct}
{\mcitedefaultendpunct}{\mcitedefaultseppunct}\relax
\EndOfBibitem
\bibitem[Ten-no(2004)]{tenno2004a}
S.~Ten-no, \emph{J. Chem. Phys.}, 2004, \textbf{121}, 117--129\relax
\mciteBstWouldAddEndPuncttrue
\mciteSetBstMidEndSepPunct{\mcitedefaultmidpunct}
{\mcitedefaultendpunct}{\mcitedefaultseppunct}\relax
\EndOfBibitem
\bibitem[Valeev(2004)]{v2004}
E.~F. Valeev, \emph{Chem. Phys. Lett.}, 2004, \textbf{395}, 190--195\relax
\mciteBstWouldAddEndPuncttrue
\mciteSetBstMidEndSepPunct{\mcitedefaultmidpunct}
{\mcitedefaultendpunct}{\mcitedefaultseppunct}\relax
\EndOfBibitem
\bibitem[Tew and Klopper(2005)]{tk2005}
D.~P. Tew and W.~Klopper, \emph{J. Chem. Phys.}, 2005, \textbf{123},
  074101\relax
\mciteBstWouldAddEndPuncttrue
\mciteSetBstMidEndSepPunct{\mcitedefaultmidpunct}
{\mcitedefaultendpunct}{\mcitedefaultseppunct}\relax
\EndOfBibitem
\bibitem[Ked\v{z}uch \emph{et~al.}(2005)Ked\v{z}uch, Milko, and
  Noga]{kedzuch:05}
S.~Ked\v{z}uch, M.~Milko and J.~Noga, \emph{Int. J. Quantum Chem.}, 2005,
  \textbf{105}, 929\relax
\mciteBstWouldAddEndPuncttrue
\mciteSetBstMidEndSepPunct{\mcitedefaultmidpunct}
{\mcitedefaultendpunct}{\mcitedefaultseppunct}\relax
\EndOfBibitem
\bibitem[Fliegl \emph{et~al.}(2005)Fliegl, Klopper, and H\"attig]{fkh05}
H.~Fliegl, W.~Klopper and C.~H\"attig, \emph{J. Chem. Phys.}, 2005,
  \textbf{122}, 084107\relax
\mciteBstWouldAddEndPuncttrue
\mciteSetBstMidEndSepPunct{\mcitedefaultmidpunct}
{\mcitedefaultendpunct}{\mcitedefaultseppunct}\relax
\EndOfBibitem
\bibitem[Fliegl \emph{et~al.}(2006)Fliegl, H\"attig, and Klopper]{fhk06b}
H.~Fliegl, C.~H\"attig and W.~Klopper, \emph{Int. J. Quantum Chem.}, 2006,
  \textbf{106}, 2306\relax
\mciteBstWouldAddEndPuncttrue
\mciteSetBstMidEndSepPunct{\mcitedefaultmidpunct}
{\mcitedefaultendpunct}{\mcitedefaultseppunct}\relax
\EndOfBibitem
\bibitem[Werner \emph{et~al.}(2007)Werner, Adler, and Manby]{f12g}
H.-J. Werner, T.~B. Adler and F.~R. Manby, \emph{J. Chem. Phys.}, 2007,
  \textbf{126}, 164102\relax
\mciteBstWouldAddEndPuncttrue
\mciteSetBstMidEndSepPunct{\mcitedefaultmidpunct}
{\mcitedefaultendpunct}{\mcitedefaultseppunct}\relax
\EndOfBibitem
\bibitem[Noga \emph{et~al.}(2007)Noga, Ked\v{z}uch, and \v{S}imunek]{noga:07}
J.~Noga, S.~Ked\v{z}uch and J.~\v{S}imunek, \emph{J. Chem. Phys.}, 2007,
  \textbf{127}, 034106\relax
\mciteBstWouldAddEndPuncttrue
\mciteSetBstMidEndSepPunct{\mcitedefaultmidpunct}
{\mcitedefaultendpunct}{\mcitedefaultseppunct}\relax
\EndOfBibitem
\bibitem[Adler \emph{et~al.}(2007)Adler, Knizia, and Werner]{ccf12}
T.~B. Adler, G.~Knizia and H.-J. Werner, \emph{J. Chem. Phys.}, 2007,
  \textbf{127}, 221106\relax
\mciteBstWouldAddEndPuncttrue
\mciteSetBstMidEndSepPunct{\mcitedefaultmidpunct}
{\mcitedefaultendpunct}{\mcitedefaultseppunct}\relax
\EndOfBibitem
\bibitem[Tew \emph{et~al.}(2007)Tew, Klopper, Neiss, and H\"attig]{tknh07}
D.~P. Tew, W.~Klopper, C.~Neiss and C.~H\"attig, \emph{Phys. Chem. Chem.
  Phys.}, 2007, \textbf{9}, 1921--1930\relax
\mciteBstWouldAddEndPuncttrue
\mciteSetBstMidEndSepPunct{\mcitedefaultmidpunct}
{\mcitedefaultendpunct}{\mcitedefaultseppunct}\relax
\EndOfBibitem
\bibitem[Knizia and Werner(2008)]{rmp2f12}
G.~Knizia and H.-J. Werner, \emph{J. Chem. Phys.}, 2008, \textbf{128},
  154103\relax
\mciteBstWouldAddEndPuncttrue
\mciteSetBstMidEndSepPunct{\mcitedefaultmidpunct}
{\mcitedefaultendpunct}{\mcitedefaultseppunct}\relax
\EndOfBibitem
\bibitem[Shiozaki \emph{et~al.}(2008)Shiozaki, Kamiya, Hirata, and
  Valeev]{skhv08a}
T.~Shiozaki, M.~Kamiya, S.~Hirata and E.~F. Valeev, \emph{J. Chem. Phys.},
  2008, \textbf{129}, 071101\relax
\mciteBstWouldAddEndPuncttrue
\mciteSetBstMidEndSepPunct{\mcitedefaultmidpunct}
{\mcitedefaultendpunct}{\mcitedefaultseppunct}\relax
\EndOfBibitem
\bibitem[Shiozaki \emph{et~al.}(2008)Shiozaki, Kamiya, Hirata, and
  Valeev]{skhv08b}
T.~Shiozaki, M.~Kamiya, S.~Hirata and E.~F. Valeev, \emph{Phys. Chem. Chem.
  Phys.}, 2008, \textbf{10}, 3358\relax
\mciteBstWouldAddEndPuncttrue
\mciteSetBstMidEndSepPunct{\mcitedefaultmidpunct}
{\mcitedefaultendpunct}{\mcitedefaultseppunct}\relax
\EndOfBibitem
\bibitem[Noga \emph{et~al.}(2008)Noga, Ked\v{z}uch, \v{S}imunek, and
  Ten-no]{nkst08}
J.~Noga, S.~Ked\v{z}uch, J.~\v{S}imunek and S.~Ten-no, \emph{J. Chem. Phys.},
  2008, \textbf{128}, 174103\relax
\mciteBstWouldAddEndPuncttrue
\mciteSetBstMidEndSepPunct{\mcitedefaultmidpunct}
{\mcitedefaultendpunct}{\mcitedefaultseppunct}\relax
\EndOfBibitem
\bibitem[Tew \emph{et~al.}(2008)Tew, Klopper, and H{\"a}ttig]{tkh2008}
D.~P. Tew, W.~Klopper and C.~H{\"a}ttig, \emph{Chem. Phys. Lett.}, 2008,
  \textbf{452}, 326--332\relax
\mciteBstWouldAddEndPuncttrue
\mciteSetBstMidEndSepPunct{\mcitedefaultmidpunct}
{\mcitedefaultendpunct}{\mcitedefaultseppunct}\relax
\EndOfBibitem
\bibitem[Valeev(2008)]{valeev08}
E.~F. Valeev, \emph{Phys. Chem. Chem. Phys.}, 2008, \textbf{10}, 106--113\relax
\mciteBstWouldAddEndPuncttrue
\mciteSetBstMidEndSepPunct{\mcitedefaultmidpunct}
{\mcitedefaultendpunct}{\mcitedefaultseppunct}\relax
\EndOfBibitem
\bibitem[Valeev and Crawford(2008)]{valeev08b}
E.~F. Valeev and T.~D. Crawford, \emph{J. Chem. Phys.}, 2008, \textbf{128},
  244113\relax
\mciteBstWouldAddEndPuncttrue
\mciteSetBstMidEndSepPunct{\mcitedefaultmidpunct}
{\mcitedefaultendpunct}{\mcitedefaultseppunct}\relax
\EndOfBibitem
\bibitem[Torheyden and Valeev(2008)]{Torheyden:2008}
M.~Torheyden and E.~F. Valeev, \emph{Phys. Chem. Chem. Phys.}, 2008,
  \textbf{10}, 3410--3420\relax
\mciteBstWouldAddEndPuncttrue
\mciteSetBstMidEndSepPunct{\mcitedefaultmidpunct}
{\mcitedefaultendpunct}{\mcitedefaultseppunct}\relax
\EndOfBibitem
\bibitem[Bokhan \emph{et~al.}(2008)Bokhan, Ten-no, and Noga]{bokhan2008}
D.~Bokhan, S.~Ten-no and J.~Noga, \emph{Phys. Chem. Chem. Phys.}, 2008,
  \textbf{10}, 3320--3326\relax
\mciteBstWouldAddEndPuncttrue
\mciteSetBstMidEndSepPunct{\mcitedefaultmidpunct}
{\mcitedefaultendpunct}{\mcitedefaultseppunct}\relax
\EndOfBibitem
\bibitem[Knizia \emph{et~al.}(2009)Knizia, Adler, and Werner]{kaw2009}
G.~Knizia, T.~B. Adler and H.-J. Werner, \emph{J. Chem. Phys.}, 2009,
  \textbf{130}, 054104\relax
\mciteBstWouldAddEndPuncttrue
\mciteSetBstMidEndSepPunct{\mcitedefaultmidpunct}
{\mcitedefaultendpunct}{\mcitedefaultseppunct}\relax
\EndOfBibitem
\bibitem[Werner \emph{et~al.}(2011)Werner, Knizia, and Manby]{Werner:2011}
H.-J. Werner, G.~Knizia and F.~R. Manby, \emph{Mol. Phys.}, 2011, \textbf{109},
  407\relax
\mciteBstWouldAddEndPuncttrue
\mciteSetBstMidEndSepPunct{\mcitedefaultmidpunct}
{\mcitedefaultendpunct}{\mcitedefaultseppunct}\relax
\EndOfBibitem
\bibitem[Ten-no(2012)]{Ten-no:TCA131-1}
S.~Ten-no, \emph{Theor. Chim. Acta}, 2012, \textbf{131}, 1070\relax
\mciteBstWouldAddEndPuncttrue
\mciteSetBstMidEndSepPunct{\mcitedefaultmidpunct}
{\mcitedefaultendpunct}{\mcitedefaultseppunct}\relax
\EndOfBibitem
\bibitem[H\"attig \emph{et~al.}(2012)H\"attig, Klopper, K\"ohn, and
  Tew]{Hattig:CR112-4}
C.~H\"attig, W.~Klopper, A.~K\"ohn and D.~P. Tew, \emph{Chem. Rev.}, 2012,
  \textbf{112}, 4--74\relax
\mciteBstWouldAddEndPuncttrue
\mciteSetBstMidEndSepPunct{\mcitedefaultmidpunct}
{\mcitedefaultendpunct}{\mcitedefaultseppunct}\relax
\EndOfBibitem
\bibitem[Kong \emph{et~al.}(2012)Kong, Bischoff, and Valeev]{Kong:CR112-75}
L.~Kong, F.~A. Bischoff and E.~F. Valeev, \emph{Chem. Rev.}, 2012,
  \textbf{112}, 75--107\relax
\mciteBstWouldAddEndPuncttrue
\mciteSetBstMidEndSepPunct{\mcitedefaultmidpunct}
{\mcitedefaultendpunct}{\mcitedefaultseppunct}\relax
\EndOfBibitem
\bibitem[Tew and Kats(2018)]{tewRelaxing2018}
D.~P. Tew and D.~Kats, \emph{J. Chem. Theory Comput.}, 2018, \textbf{14},
  5435\relax
\mciteBstWouldAddEndPuncttrue
\mciteSetBstMidEndSepPunct{\mcitedefaultmidpunct}
{\mcitedefaultendpunct}{\mcitedefaultseppunct}\relax
\EndOfBibitem
\bibitem[Kats and Tew(2019)]{katsOrbitalOptimized2019}
D.~Kats and D.~P. Tew, \emph{J. Chem. Theory Comput.}, 2019, \textbf{15},
  13\relax
\mciteBstWouldAddEndPuncttrue
\mciteSetBstMidEndSepPunct{\mcitedefaultmidpunct}
{\mcitedefaultendpunct}{\mcitedefaultseppunct}\relax
\EndOfBibitem
\bibitem[Shiozaki and Werner(2013)]{shiozaki_multireference_2013}
T.~Shiozaki and H.-J. Werner, \emph{Mol. Phys.}, 2013, \textbf{111}, 607\relax
\mciteBstWouldAddEndPuncttrue
\mciteSetBstMidEndSepPunct{\mcitedefaultmidpunct}
{\mcitedefaultendpunct}{\mcitedefaultseppunct}\relax
\EndOfBibitem
\bibitem[Booth \emph{et~al.}(2012)Booth, Cleland, Alavi, and
  Tew]{boothExplicitly2012}
G.~H. Booth, D.~Cleland, A.~Alavi and D.~P. Tew, \emph{J. Chem. Phys.}, 2012,
  \textbf{137}, 164112\relax
\mciteBstWouldAddEndPuncttrue
\mciteSetBstMidEndSepPunct{\mcitedefaultmidpunct}
{\mcitedefaultendpunct}{\mcitedefaultseppunct}\relax
\EndOfBibitem
\bibitem[Manby \emph{et~al.}(2006)Manby, Werner, Adler, and May]{Manby:06a}
F.~R. Manby, H.-J. Werner, T.~B. Adler and A.~J. May, \emph{J. Chem. Phys.},
  2006, \textbf{124}, 094103\relax
\mciteBstWouldAddEndPuncttrue
\mciteSetBstMidEndSepPunct{\mcitedefaultmidpunct}
{\mcitedefaultendpunct}{\mcitedefaultseppunct}\relax
\EndOfBibitem
\bibitem[Tew \emph{et~al.}(2011)Tew, Helmich, and H\"attig]{Tew:2011}
D.~P. Tew, B.~Helmich and C.~H\"attig, \emph{J. Chem. Phys.}, 2011,
  \textbf{135}, 074107\relax
\mciteBstWouldAddEndPuncttrue
\mciteSetBstMidEndSepPunct{\mcitedefaultmidpunct}
{\mcitedefaultendpunct}{\mcitedefaultseppunct}\relax
\EndOfBibitem
\bibitem[Ma \emph{et~al.}(2017)Ma, Schwilk, K\"{o}ppl, and Werner]{Ma:2017}
Q.~Ma, M.~Schwilk, C.~K\"{o}ppl and H.-J. Werner, \emph{J. Chem. Theory
  Comput.}, 2017, \textbf{13}, 4871--4896\relax
\mciteBstWouldAddEndPuncttrue
\mciteSetBstMidEndSepPunct{\mcitedefaultmidpunct}
{\mcitedefaultendpunct}{\mcitedefaultseppunct}\relax
\EndOfBibitem
\bibitem[Pavo\v{s}evi\'{c} \emph{et~al.}(2017)Pavo\v{s}evi\'{c}, Peng, Pinski,
  Riplinger, Neese, and Valeev]{Pavosevic:2017}
F.~Pavo\v{s}evi\'{c}, C.~Peng, P.~Pinski, C.~Riplinger, F.~Neese and E.~F.
  Valeev, \emph{J. Chem. Phys.}, 2017, \textbf{146}, 174108\relax
\mciteBstWouldAddEndPuncttrue
\mciteSetBstMidEndSepPunct{\mcitedefaultmidpunct}
{\mcitedefaultendpunct}{\mcitedefaultseppunct}\relax
\EndOfBibitem
\bibitem[Usvyat(2013)]{usvyat_linear-scaling_2013}
D.~Usvyat, \emph{J. Chem. Phys.}, 2013, \textbf{139}, 194101\relax
\mciteBstWouldAddEndPuncttrue
\mciteSetBstMidEndSepPunct{\mcitedefaultmidpunct}
{\mcitedefaultendpunct}{\mcitedefaultseppunct}\relax
\EndOfBibitem
\bibitem[Gr{\"u}neis(2015)]{gruneisEfficient2015}
A.~Gr{\"u}neis, \emph{Phys. Rev. Lett.}, 2015, \textbf{115}, 066402\relax
\mciteBstWouldAddEndPuncttrue
\mciteSetBstMidEndSepPunct{\mcitedefaultmidpunct}
{\mcitedefaultendpunct}{\mcitedefaultseppunct}\relax
\EndOfBibitem
\bibitem[K{\"o}hn(2009)]{Koehn:2009}
A.~K{\"o}hn, \emph{J. Chem. Phys.}, 2009, \textbf{130}, 131101\relax
\mciteBstWouldAddEndPuncttrue
\mciteSetBstMidEndSepPunct{\mcitedefaultmidpunct}
{\mcitedefaultendpunct}{\mcitedefaultseppunct}\relax
\EndOfBibitem
\bibitem[Boys and Handy(1969)]{boysCalculation1969}
S.~F. Boys and N.~C. Handy, \emph{Proc. R. Soc. A}, 1969, \textbf{310},
  63--78\relax
\mciteBstWouldAddEndPuncttrue
\mciteSetBstMidEndSepPunct{\mcitedefaultmidpunct}
{\mcitedefaultendpunct}{\mcitedefaultseppunct}\relax
\EndOfBibitem
\bibitem[{Ten-no}(2000)]{ten-noFeasible2000}
S.~{Ten-no}, \emph{Chem. Phys. Lett.}, 2000, \textbf{330}, 169--174\relax
\mciteBstWouldAddEndPuncttrue
\mciteSetBstMidEndSepPunct{\mcitedefaultmidpunct}
{\mcitedefaultendpunct}{\mcitedefaultseppunct}\relax
\EndOfBibitem
\bibitem[Hino \emph{et~al.}(2001)Hino, Tanimura, and Ten-no]{htt2001}
O.~Hino, Y.~Tanimura and S.~Ten-no, \emph{J. Chem. Phys.}, 2001, \textbf{115},
  7865--7871\relax
\mciteBstWouldAddEndPuncttrue
\mciteSetBstMidEndSepPunct{\mcitedefaultmidpunct}
{\mcitedefaultendpunct}{\mcitedefaultseppunct}\relax
\EndOfBibitem
\bibitem[Hino \emph{et~al.}(2002)Hino, Tanimura, and
  {Ten-no}]{hinoApplication2002}
O.~Hino, Y.~Tanimura and S.~{Ten-no}, \emph{Chem. Phys. Lett.}, 2002,
  \textbf{353}, 317--323\relax
\mciteBstWouldAddEndPuncttrue
\mciteSetBstMidEndSepPunct{\mcitedefaultmidpunct}
{\mcitedefaultendpunct}{\mcitedefaultseppunct}\relax
\EndOfBibitem
\bibitem[Umezawa and Tsuneyuki(2003)]{umezawa_transcorrelated_2003}
N.~Umezawa and S.~Tsuneyuki, \emph{J. Chem. Phys.}, 2003, \textbf{119},
  10015--10031\relax
\mciteBstWouldAddEndPuncttrue
\mciteSetBstMidEndSepPunct{\mcitedefaultmidpunct}
{\mcitedefaultendpunct}{\mcitedefaultseppunct}\relax
\EndOfBibitem
\bibitem[Yanai and Chan(2006)]{yanai_canonical_2006}
T.~Yanai and G.~K.-L. Chan, \emph{J. Chem. Phys.}, 2006, \textbf{124},
  194106\relax
\mciteBstWouldAddEndPuncttrue
\mciteSetBstMidEndSepPunct{\mcitedefaultmidpunct}
{\mcitedefaultendpunct}{\mcitedefaultseppunct}\relax
\EndOfBibitem
\bibitem[Yanai and Chan(2007)]{yanai_canonical_2007}
T.~Yanai and G.~K.-L. Chan, \emph{J. Chem. Phys.}, 2007, \textbf{127},
  104107\relax
\mciteBstWouldAddEndPuncttrue
\mciteSetBstMidEndSepPunct{\mcitedefaultmidpunct}
{\mcitedefaultendpunct}{\mcitedefaultseppunct}\relax
\EndOfBibitem
\bibitem[Yanai and Shiozaki(2012)]{yanai_canonical_2012}
T.~Yanai and T.~Shiozaki, \emph{J. Chem. Phys.}, 2012, \textbf{136},
  084107\relax
\mciteBstWouldAddEndPuncttrue
\mciteSetBstMidEndSepPunct{\mcitedefaultmidpunct}
{\mcitedefaultendpunct}{\mcitedefaultseppunct}\relax
\EndOfBibitem
\bibitem[Tsuneyuki(2008)]{tsuneyukiTranscorrelated2008}
S.~Tsuneyuki, \emph{Prog. Theor. Phys. Suppl.}, 2008, \textbf{176},
  134--142\relax
\mciteBstWouldAddEndPuncttrue
\mciteSetBstMidEndSepPunct{\mcitedefaultmidpunct}
{\mcitedefaultendpunct}{\mcitedefaultseppunct}\relax
\EndOfBibitem
\bibitem[Ochi \emph{et~al.}(2012)Ochi, Sodeyama, Sakuma, and
  Tsuneyuki]{ochi_efficient_2012}
M.~Ochi, K.~Sodeyama, R.~Sakuma and S.~Tsuneyuki, \emph{J. Chem. Phys.}, 2012,
  \textbf{136}, 094108\relax
\mciteBstWouldAddEndPuncttrue
\mciteSetBstMidEndSepPunct{\mcitedefaultmidpunct}
{\mcitedefaultendpunct}{\mcitedefaultseppunct}\relax
\EndOfBibitem
\bibitem[Ochi and Tsuneyuki(2014)]{ochi_optical_2014}
M.~Ochi and S.~Tsuneyuki, \emph{J. Chem. Theory Comput.}, 2014, \textbf{10},
  4098--4103\relax
\mciteBstWouldAddEndPuncttrue
\mciteSetBstMidEndSepPunct{\mcitedefaultmidpunct}
{\mcitedefaultendpunct}{\mcitedefaultseppunct}\relax
\EndOfBibitem
\bibitem[Ochi and Tsuneyuki(2015)]{ochi_second-order_2015}
M.~Ochi and S.~Tsuneyuki, \emph{Chem. Phys. Lett.}, 2015, \textbf{621},
  177--183\relax
\mciteBstWouldAddEndPuncttrue
\mciteSetBstMidEndSepPunct{\mcitedefaultmidpunct}
{\mcitedefaultendpunct}{\mcitedefaultseppunct}\relax
\EndOfBibitem
\bibitem[Ochi \emph{et~al.}(2016)Ochi, Yamamoto, Arita, and
  Tsuneyuki]{ochi_iterative_2016}
M.~Ochi, Y.~Yamamoto, R.~Arita and S.~Tsuneyuki, \emph{J. Chem. Phys.}, 2016,
  \textbf{144}, 104109\relax
\mciteBstWouldAddEndPuncttrue
\mciteSetBstMidEndSepPunct{\mcitedefaultmidpunct}
{\mcitedefaultendpunct}{\mcitedefaultseppunct}\relax
\EndOfBibitem
\bibitem[Wahlen-Strothman \emph{et~al.}(2015)Wahlen-Strothman,
  Jim{\'e}nez-Hoyos, Henderson, and Scuseria]{wahlen-strothman_lie_2015}
J.~M. Wahlen-Strothman, C.~A. Jim{\'e}nez-Hoyos, T.~M. Henderson and G.~E.
  Scuseria, \emph{Phys. Rev. B}, 2015, \textbf{91}, 041114\relax
\mciteBstWouldAddEndPuncttrue
\mciteSetBstMidEndSepPunct{\mcitedefaultmidpunct}
{\mcitedefaultendpunct}{\mcitedefaultseppunct}\relax
\EndOfBibitem
\bibitem[Luo and Alavi(2018)]{luoCombining2018}
H.~Luo and A.~Alavi, \emph{J. Chem. Theory Comput.}, 2018, \textbf{14},
  1403--1411\relax
\mciteBstWouldAddEndPuncttrue
\mciteSetBstMidEndSepPunct{\mcitedefaultmidpunct}
{\mcitedefaultendpunct}{\mcitedefaultseppunct}\relax
\EndOfBibitem
\bibitem[Dobrautz \emph{et~al.}(2019)Dobrautz, Luo, and
  Alavi]{dobrautzCompact2019}
W.~Dobrautz, H.~Luo and A.~Alavi, \emph{Phys. Rev. B}, 2019, \textbf{99},
  075119\relax
\mciteBstWouldAddEndPuncttrue
\mciteSetBstMidEndSepPunct{\mcitedefaultmidpunct}
{\mcitedefaultendpunct}{\mcitedefaultseppunct}\relax
\EndOfBibitem
\bibitem[Cohen \emph{et~al.}(2019)Cohen, Luo, Guther, Dobrautz, Tew, and
  Alavi]{cohenSimilarity2019}
A.~J. Cohen, H.~Luo, K.~Guther, W.~Dobrautz, D.~P. Tew and A.~Alavi, \emph{J.
  Chem. Phys.}, 2019, \textbf{151}, 061101\relax
\mciteBstWouldAddEndPuncttrue
\mciteSetBstMidEndSepPunct{\mcitedefaultmidpunct}
{\mcitedefaultendpunct}{\mcitedefaultseppunct}\relax
\EndOfBibitem
\bibitem[Baiardi and Reiher(2020)]{baiardi_transcorrelated_2020}
A.~Baiardi and M.~Reiher, \emph{J. Chem. Phys.}, 2020, \textbf{153},
  164115\relax
\mciteBstWouldAddEndPuncttrue
\mciteSetBstMidEndSepPunct{\mcitedefaultmidpunct}
{\mcitedefaultendpunct}{\mcitedefaultseppunct}\relax
\EndOfBibitem
\bibitem[Khamoshi \emph{et~al.}(2021)Khamoshi, Chen, Henderson, and
  Scuseria]{khamoshi_exploring_2021}
A.~Khamoshi, G.~P. Chen, T.~M. Henderson and G.~E. Scuseria, \emph{J. Chem.
  Phys.}, 2021, \textbf{154}, 074113\relax
\mciteBstWouldAddEndPuncttrue
\mciteSetBstMidEndSepPunct{\mcitedefaultmidpunct}
{\mcitedefaultendpunct}{\mcitedefaultseppunct}\relax
\EndOfBibitem
\bibitem[Giner(2021)]{giner_new_2021}
E.~Giner, \emph{J. Chem. Phys.}, 2021, \textbf{154}, 084119\relax
\mciteBstWouldAddEndPuncttrue
\mciteSetBstMidEndSepPunct{\mcitedefaultmidpunct}
{\mcitedefaultendpunct}{\mcitedefaultseppunct}\relax
\EndOfBibitem
\bibitem[Guther \emph{et~al.}(2021)Guther, Cohen, Luo, and
  Alavi]{gutherberyliumdimer2021}
K.~Guther, A.~J. Cohen, H.~Luo and A.~Alavi, \emph{The Journal of Chemical
  Physics}, 2021, \textbf{155}, 011102\relax
\mciteBstWouldAddEndPuncttrue
\mciteSetBstMidEndSepPunct{\mcitedefaultmidpunct}
{\mcitedefaultendpunct}{\mcitedefaultseppunct}\relax
\EndOfBibitem
\bibitem[Liao \emph{et~al.}(2021)Liao, Schraivogel, Luo, Kats, and
  Alavi]{liaoEfficient2021}
K.~Liao, T.~Schraivogel, H.~Luo, D.~Kats and A.~Alavi, \emph{Phys. Rev.
  Research}, 2021, \textbf{3}, 033072\relax
\mciteBstWouldAddEndPuncttrue
\mciteSetBstMidEndSepPunct{\mcitedefaultmidpunct}
{\mcitedefaultendpunct}{\mcitedefaultseppunct}\relax
\EndOfBibitem
\bibitem[Liao \emph{et~al.}(2023)Liao, Zhai, Christlmaier, Schraivogel, R\'ios,
  Kats, and Alavi]{liao22}
K.~Liao, H.~Zhai, E.~M. Christlmaier, T.~Schraivogel, P.~L. R\'ios, D.~Kats and
  A.~Alavi, \emph{J. Chem. Theory Comput.}, 2023, \textbf{19}, 1734\relax
\mciteBstWouldAddEndPuncttrue
\mciteSetBstMidEndSepPunct{\mcitedefaultmidpunct}
{\mcitedefaultendpunct}{\mcitedefaultseppunct}\relax
\EndOfBibitem
\bibitem[Haupt \emph{et~al.}(2023)Haupt, Hosseini, L{\'o}pez~R{\'i}os,
  Dobrautz, Cohen, and Alavi]{hauptOptimizing2023}
J.~P. Haupt, S.~M. Hosseini, P.~L{\'o}pez~R{\'i}os, W.~Dobrautz, A.~Cohen and
  A.~Alavi, \emph{J. Chem. Phys.}, 2023, \textbf{158}, 224105\relax
\mciteBstWouldAddEndPuncttrue
\mciteSetBstMidEndSepPunct{\mcitedefaultmidpunct}
{\mcitedefaultendpunct}{\mcitedefaultseppunct}\relax
\EndOfBibitem
\bibitem[Christlmaier \emph{et~al.}(2023)Christlmaier, Schraivogel,
  L{\'o}pez~R{\'i}os, Alavi, and Kats]{christlmaierXTC2023}
E.~M. Christlmaier, T.~Schraivogel, P.~L{\'o}pez~R{\'i}os, A.~Alavi and
  D.~Kats, \emph{J. Chem. Phys.}, 2023, \textbf{159}, 014113\relax
\mciteBstWouldAddEndPuncttrue
\mciteSetBstMidEndSepPunct{\mcitedefaultmidpunct}
{\mcitedefaultendpunct}{\mcitedefaultseppunct}\relax
\EndOfBibitem
\bibitem[Ammar \emph{et~al.}(2023)Ammar, Scemama, and
  Giner]{ammarBiorthonormal2023}
A.~Ammar, A.~Scemama and E.~Giner, \emph{J. Chem. Theory Comput.}, 2023,
  \textbf{19}, 4883\relax
\mciteBstWouldAddEndPuncttrue
\mciteSetBstMidEndSepPunct{\mcitedefaultmidpunct}
{\mcitedefaultendpunct}{\mcitedefaultseppunct}\relax
\EndOfBibitem
\bibitem[Lee and Thom(2023)]{leeStudies2023}
N.~Lee and A.~J.~W. Thom, \emph{J. Chem. Theory Comput.}, 2023, \textbf{19},
  5743\relax
\mciteBstWouldAddEndPuncttrue
\mciteSetBstMidEndSepPunct{\mcitedefaultmidpunct}
{\mcitedefaultendpunct}{\mcitedefaultseppunct}\relax
\EndOfBibitem
\bibitem[Ammar \emph{et~al.}(2023)Ammar, Scemama, and
  Giner]{ammarTranscorrelated2023}
A.~Ammar, A.~Scemama and E.~Giner, \emph{J. Chem. Phys.}, 2023, \textbf{159},
  114121\relax
\mciteBstWouldAddEndPuncttrue
\mciteSetBstMidEndSepPunct{\mcitedefaultmidpunct}
{\mcitedefaultendpunct}{\mcitedefaultseppunct}\relax
\EndOfBibitem
\bibitem[Schraivogel \emph{et~al.}(2023)Schraivogel, Christlmaier,
  L{\'o}pez~R{\'i}os, Alavi, and Kats]{schraivogelTranscorrelated2023}
T.~Schraivogel, E.~M. Christlmaier, P.~L{\'o}pez~R{\'i}os, A.~Alavi and
  D.~Kats, \emph{J. Chem. Phys.}, 2023, \textbf{158}, 214106\relax
\mciteBstWouldAddEndPuncttrue
\mciteSetBstMidEndSepPunct{\mcitedefaultmidpunct}
{\mcitedefaultendpunct}{\mcitedefaultseppunct}\relax
\EndOfBibitem
\bibitem[Schraivogel \emph{et~al.}(2021)Schraivogel, Cohen, Alavi, and
  Kats]{TCCC2021}
T.~Schraivogel, A.~J. Cohen, A.~Alavi and D.~Kats, \emph{J. Chem. Phys.}, 2021,
  \textbf{155}, 191101\relax
\mciteBstWouldAddEndPuncttrue
\mciteSetBstMidEndSepPunct{\mcitedefaultmidpunct}
{\mcitedefaultendpunct}{\mcitedefaultseppunct}\relax
\EndOfBibitem
\bibitem[Kats \emph{et~al.}(2024)Kats, Schraivogel, Hauskrecht, Rickert, and
  Wu]{elemcojl}
D.~Kats, T.~Schraivogel, J.~Hauskrecht, C.~Rickert and F.~Wu,
  \emph{\texttt{ElemCo.jl}: Julia program package for electron correlation
  methods}, 2024, see github.com/fkfest/ElemCo.jl\relax
\mciteBstWouldAddEndPuncttrue
\mciteSetBstMidEndSepPunct{\mcitedefaultmidpunct}
{\mcitedefaultendpunct}{\mcitedefaultseppunct}\relax
\EndOfBibitem
\bibitem[ele()]{elemcojl-docs}
\emph{\texttt{ElemCo.jl} documentation}, \url{https://elem.co.il}, accessed
  2024-02-25\relax
\mciteBstWouldAddEndPuncttrue
\mciteSetBstMidEndSepPunct{\mcitedefaultmidpunct}
{\mcitedefaultendpunct}{\mcitedefaultseppunct}\relax
\EndOfBibitem
\bibitem[Taube and Bartlett(2008)]{taubeImproving2008}
A.~G. Taube and R.~J. Bartlett, \emph{J. Chem. Phys.}, 2008, \textbf{128},
  044110\relax
\mciteBstWouldAddEndPuncttrue
\mciteSetBstMidEndSepPunct{\mcitedefaultmidpunct}
{\mcitedefaultendpunct}{\mcitedefaultseppunct}\relax
\EndOfBibitem
\bibitem[Drummond \emph{et~al.}(2004)Drummond, Towler, and
  Needs]{drummond_jastrow_2004}
N.~D. Drummond, M.~D. Towler and R.~J. Needs, \emph{Phys. Rev. B}, 2004,
  \textbf{70}, 235119\relax
\mciteBstWouldAddEndPuncttrue
\mciteSetBstMidEndSepPunct{\mcitedefaultmidpunct}
{\mcitedefaultendpunct}{\mcitedefaultseppunct}\relax
\EndOfBibitem
\bibitem[Needs \emph{et~al.}(2020)Needs, Towler, Drummond, L\'opez~R\'ios, and
  Trail]{needs20}
R.~J. Needs, M.~D. Towler, N.~D. Drummond, P.~L\'opez~R\'ios and J.~R. Trail,
  \emph{J. Chem. Phys.}, 2020, \textbf{152}, 154106\relax
\mciteBstWouldAddEndPuncttrue
\mciteSetBstMidEndSepPunct{\mcitedefaultmidpunct}
{\mcitedefaultendpunct}{\mcitedefaultseppunct}\relax
\EndOfBibitem
\bibitem[Haupt \emph{et~al.}(2023)Haupt, L{\'o}pez~R{\'i}os, Christlmaier,
  Hauskrecht, Liao, Dobrautz, Guther, Cohen, and Alavi]{tchint}
J.~P. Haupt, P.~L{\'o}pez~R{\'i}os, E.~M. Christlmaier, J.~Hauskrecht, K.~Liao,
  W.~Dobrautz, K.~Guther, A.~J. Cohen and A.~Alavi, \emph{\texttt{TCHINT}:
  Transcorrelated Hamiltonian integral library}, 2023, to be released\relax
\mciteBstWouldAddEndPuncttrue
\mciteSetBstMidEndSepPunct{\mcitedefaultmidpunct}
{\mcitedefaultendpunct}{\mcitedefaultseppunct}\relax
\EndOfBibitem
\bibitem[Werner \emph{et~al.}()Werner, Knowles, Celani, Gy\"orffy, Hesselmann,
  Kats, Knizia, K\"ohn, Korona, Kreplin, Lindh, Ma, Manby, Mitrushenkov,
  Rauhut, {Sch\"{u}tz}, Shamasundar, Adler, Amos, Bennie, Bernhardsson,
  Berning, Black, Bygrave, Cimiraglia, Cooper, Coughtrie, Deegan, Dobbyn, Doll,
  Dornbach, Eckert, Erfort, Goll, Hampel, Hetzer, Hill, Hodges, Hrenar, Jansen,
  K\"oppl, Kollmar, Lee, Liu, Lloyd, Mata, May, Mussard, McNicholas, Meyer,
  {Miller III}, Mura, Nicklass, O'Neill, Palmieri, Peng, Peterson, Pfl\"uger,
  Pitzer, Polyak, Reiher, Richardson, Robinson, Schr\"oder, Schwilk, Shiozaki,
  Sibaev, Stoll, Stone, Tarroni, Thorsteinsson, Toulouse, Wang, Welborn, and
  Ziegler]{MOLPRO}
H.-J. Werner, P.~J. Knowles, P.~Celani, W.~Gy\"orffy, A.~Hesselmann, D.~Kats,
  G.~Knizia, A.~K\"ohn, T.~Korona, D.~Kreplin, R.~Lindh, Q.~Ma, F.~R. Manby,
  A.~Mitrushenkov, G.~Rauhut, M.~{Sch\"{u}tz}, K.~R. Shamasundar, T.~B. Adler,
  R.~D. Amos, S.~J. Bennie, A.~Bernhardsson, A.~Berning, J.~A. Black, P.~J.
  Bygrave, R.~Cimiraglia, D.~L. Cooper, D.~Coughtrie, M.~J.~O. Deegan, A.~J.
  Dobbyn, K.~Doll, M.~Dornbach, F.~Eckert, S.~Erfort, E.~Goll, C.~Hampel,
  G.~Hetzer, J.~G. Hill, M.~Hodges, T.~Hrenar, G.~Jansen, C.~K\"oppl,
  C.~Kollmar, S.~J.~R. Lee, Y.~Liu, A.~W. Lloyd, R.~A. Mata, A.~J. May,
  B.~Mussard, S.~J. McNicholas, W.~Meyer, T.~F. {Miller III}, M.~E. Mura,
  A.~Nicklass, D.~P. O'Neill, P.~Palmieri, D.~Peng, K.~A. Peterson,
  K.~Pfl\"uger, R.~Pitzer, I.~Polyak, M.~Reiher, J.~O. Richardson, J.~B.
  Robinson, B.~Schr\"oder, M.~Schwilk, T.~Shiozaki, M.~Sibaev, H.~Stoll, A.~J.
  Stone, R.~Tarroni, T.~Thorsteinsson, J.~Toulouse, M.~Wang, M.~Welborn and
  B.~Ziegler, \emph{MOLPRO, 2023.1 , a package of ab initio programs}, see
  https://www.molpro.net\relax
\mciteBstWouldAddEndPuncttrue
\mciteSetBstMidEndSepPunct{\mcitedefaultmidpunct}
{\mcitedefaultendpunct}{\mcitedefaultseppunct}\relax
\EndOfBibitem
\bibitem[Sun \emph{et~al.}(2018)Sun, Berkelbach, Blunt, Booth, Guo, Li, Liu,
  McClain, Sayfutyarova, Sharma, Wouters, and Chan]{PySCF}
Q.~Sun, T.~C. Berkelbach, N.~S. Blunt, G.~H. Booth, S.~Guo, Z.~Li, J.~Liu,
  J.~D. McClain, E.~R. Sayfutyarova, S.~Sharma, S.~Wouters and G.~K.-L. Chan,
  \emph{WIREs Comput. Mol. Sci.}, 2018, \textbf{8}, e1340\relax
\mciteBstWouldAddEndPuncttrue
\mciteSetBstMidEndSepPunct{\mcitedefaultmidpunct}
{\mcitedefaultendpunct}{\mcitedefaultseppunct}\relax
\EndOfBibitem
\bibitem[Tajti \emph{et~al.}(2004)Tajti, Szalay, Császár, Kállay, Gauss,
  Valeev, Flowers, Vázquez, and Stanton]{HEAT1}
A.~Tajti, P.~G. Szalay, A.~G. Császár, M.~Kállay, J.~Gauss, E.~F. Valeev,
  B.~A. Flowers, J.~Vázquez and J.~F. Stanton, \emph{J. Chem. Phys.}, 2004,
  \textbf{121}, 11599\relax
\mciteBstWouldAddEndPuncttrue
\mciteSetBstMidEndSepPunct{\mcitedefaultmidpunct}
{\mcitedefaultendpunct}{\mcitedefaultseppunct}\relax
\EndOfBibitem
\bibitem[Bomble \emph{et~al.}(2006)Bomble, Vázquez, Kállay, Michauk, Szalay,
  Császár, Gauss, and Stanton]{HEAT2}
Y.~J. Bomble, J.~Vázquez, M.~Kállay, C.~Michauk, P.~G. Szalay, A.~G.
  Császár, J.~Gauss and J.~F. Stanton, \emph{J. Chem. Phys.}, 2006,
  \textbf{125}, 064108\relax
\mciteBstWouldAddEndPuncttrue
\mciteSetBstMidEndSepPunct{\mcitedefaultmidpunct}
{\mcitedefaultendpunct}{\mcitedefaultseppunct}\relax
\EndOfBibitem
\bibitem[Harding \emph{et~al.}(2008)Harding, Vázquez, Ruscic, Wilson, Gauss,
  and Stanton]{HEAT3}
M.~E. Harding, J.~Vázquez, B.~Ruscic, A.~K. Wilson, J.~Gauss and J.~F.
  Stanton, \emph{J. Chem. Phys.}, 2008, \textbf{128}, 114111\relax
\mciteBstWouldAddEndPuncttrue
\mciteSetBstMidEndSepPunct{\mcitedefaultmidpunct}
{\mcitedefaultendpunct}{\mcitedefaultseppunct}\relax
\EndOfBibitem
\bibitem[Thorpe \emph{et~al.}(2019)Thorpe, Lopez, Nguyen, Baraban, Bross,
  Ruscic, and Stanton]{HEAT4}
J.~H. Thorpe, C.~A. Lopez, T.~L. Nguyen, J.~H. Baraban, D.~H. Bross, B.~Ruscic
  and J.~F. Stanton, \emph{J. Chem. Phys.}, 2019, \textbf{150}, 224102\relax
\mciteBstWouldAddEndPuncttrue
\mciteSetBstMidEndSepPunct{\mcitedefaultmidpunct}
{\mcitedefaultendpunct}{\mcitedefaultseppunct}\relax
\EndOfBibitem
\end{mcitethebibliography}
\bibliographystyle{rsc} 

\end{document}